\documentclass[prc,notitlepage,twocolumn,showpacs,floatfix,nofootinbib,preprintnumbers,superscriptaddress,amsmath,amssymb]{revtex4-1}
\usepackage[T1]{fontenc} 
\usepackage{amsmath}
\usepackage{amssymb}
\usepackage{graphicx}
\usepackage{hyperref}

\begin{document}

\title{Alpha Decay Energies of Superheavy Nuclei: Systematic Trends}

\author{E. Olsen}
\affiliation{Department of Physics and Astronomy and FRIB Laboratory, Michigan State University, East Lansing, Michigan 48824, USA}

\author{W. Nazarewicz}
\affiliation{Department of Physics and Astronomy and FRIB Laboratory, Michigan State University, East Lansing, Michigan 48824, USA}

\date{\today{}}

\begin{abstract}
  \begin{description}
    \item[Background] 
    New superheavy nuclei are often identified through their characteristic $\alpha$-decay energies, which requires accurate calculations of $Q_{\alpha}$ values. While many $Q_{\alpha}$ predictions are available, little is known about their uncertainties, and this makes it difficult to carry out extrapolations to yet-unknown systems.
    \item[Purpose] 
    This work aims to analyze several models, compare their predictions to available experimental data, and study their performance for the unobserved $\alpha$-decay chains of $^{296}$120 and $^{298}$120, which are of current experimental interest. Our quantified results will also serve as a benchmark for future, more sophisticated statistical studies. 
    \item[Methods] 
   We use nuclear superfluid Density Functional Theory (DFT) with several Skyrme energy density functionals (EDFs). To estimate systematic model uncertainties we employ uniform model averaging.
    \item[Results] 
    We evaluated the $Q_{\alpha}$ values for even-even nuclei from Fm to $Z=120$. For well deformed nuclei between Fm and Ds, we find excellent consistency between different model predictions, and a good agreement with experiment. For transitional nuclei beyond Ds, inter-model differences grow, resulting in an appreciable systematic error. In particular, our models underestimate $Q_{\alpha}$ for the heaviest nucleus $^{294}$Og.
    \item[Conclusions]
    The robustness of DFT predictions for well deformed superheavy nuclei supports the idea of using experimental $Q_{\alpha}$ values, together with theoretical predictions, as reasonable $(Z,A)$ indicators. Unfortunately, this identification method is not expected to work well in the region of deformed-to-spherical shape transition as one approaches $N=184$. The use of $Q_{\alpha}$ values in the identification of new superheavy nuclei will benefit greatly from both progress in developing new spectroscopic-quality EDFs and more sophisticated statistical techniques of uncertainty quantification.
  \end{description}
\end{abstract}

\maketitle


\section{Introduction}
Superheavy nuclei with $Z \geq 104$ occupy the upper right-hand corner of the nuclear chart \cite{Hoffman2000b, Seaborg1990}. The study of these massive systems has been prompted by a desire to answer many fundamental questions pertaining to nuclear and atomic physics, and chemistry \cite{Giuliani2018, Nazarewicz2018}.

In particular, the search for long-lived superheavy nuclei in nature has been active for many decades. Early theoretical calculations predicted the superheavy magic numbers (the so-called ``island of stability'' \cite{Viola1966}) at $Z=114$ and $N=184$ \cite{Myers1966, Nilsson1969, Sobiczewski1966}. As time progressed and models improved, the superheavy magic numbers were suggested at $114$, $120$, $124$, or $126$ for protons and either $172$ or $184$ for neutrons \cite{Afanasjev2003, Afanasjev2006, Bender1999, Cwiok1996, Kruppa2000, Rutz1997}. However, unlike with traditional magic numbers, these predictions for superheavy nuclei are more likely to correspond to extended half-lives rather than stable systems \cite{Dullmann2018}; this is due to both large Coulomb repulsion and the high density of single-particle levels \cite{Agbemava2015, Bender2001} resulting in a diffused shell structure \cite{Agbemava2015, Bender2001, Jerabek2018}.

Through the experimental techniques of cold and hot heavy-ion fusion \cite{Oganessian1975, Oganessian2007}, many isotopes of new elements between $Z=114$ (Fl) and $Z=118$ (Og) were discovered and added to the nuclear chart in the last decade \cite{Barber2011, Karol2016a, Karol2016b}. At present, efforts to identify nuclei beyond Og and more neutron-rich systems have been unsuccessful \cite{Dullmann2011, Dullmann2017, Oganessian2009}.

Known superheavy nuclei primarily decay through $\alpha$ emission and spontaneous fission \cite{Hofmann2004, Oganessian2015, Oganessian2017}. As a result, new isotopes are often identified through the observation of their characteristic $\alpha$-decay chains \cite{Oganessian2006} based on experimental data and theoretical predictions. As such, calculations of $Q_{\alpha}$ values with quantified uncertainties are useful for future superheavy nuclei searches.

Numerous calculations of $Q_{\alpha}$ values for superheavy nuclei are available, see, e.g., Refs.~\cite{Agbemava2015, Bender2000, Berger2004, Cwiok1999, Erler2012a, Gambhir2005, Heenen2015, Jachimowicz2014, Muntian2003, Sobiczewski2007, Tolokonnikov2013, Tolokonnikov2017, Typel2003, Warda2012}. Some of these studies also include calculations of $\alpha$-decay half-lives using empirical formulas \cite{Brown1992, Budaca2016, Chowdhury2008, Dong2011, Koura2012, Parkhomenko2005, Royer2008, Viola1966, Ward2015}, in which half-lives are expressed as functions of $Q_{\alpha}$. In this respect,  $Q_{\alpha}$ values and half-lives carry the same information content. Except for the recent surveys \cite{Agbemava2015, Heenen2015}, the emphasis of theoretical studies was on the performance of a specific model. It is the purpose of this paper to take another approach: analyze and compare $Q_{\alpha}$ values predicted by several Skyrme-DFT models. In this way, their systematic uncertainties and robustness can be estimated more thoroughly through both direct analysis of different parameterizations and model mixing.

This paper is structured as follows. In Sec.~\ref{approach} we discuss the theoretical approach used. Section~\ref{results} displays our results for known superheavy nuclei found and unknown $^{296}$120 and $^{298}$120. It also presents results of the model-mixing analysis. Finally, in Sec.~\ref{conclusions} we discuss conclusions and perspectives.

\section{Theoretical Approach}\label{approach}

All of our calculations were performed within the framework of nuclear Density Functional Theory (DFT) \cite{Bender2003}, where the total energy of the system is expressed in terms of the energy density, which is a functional of one-body local densities and currents. Nuclear DFT is the tool of choice for making global predictions for complex heavy nuclei. As emphasized in Ref.~\cite{Stoitsov2006}, this method is general enough to be applied anywhere on the nuclear chart; it can incorporate nuclear deformations through intrinsic symmetry breaking; and it can provide quantified predictions for a variety of observables. The main ingredient of nuclear DFT is the energy density functional (EDF), which represents an effective in-medium nuclear interaction. An EDF contains a number of coupling constants which are adjusted to selected experimental data and theoretical pseudo-data \cite{Bender2003, Klupfel2009, Kortelainen2010}; depending on the optimization methodology and strategy chosen, these low-energy couplings change, and a new EDF parameterization is developed.

For this work, seven effective Skyrme EDFs \cite{Skyrme1958, Vautherin1972} in the particle-hole channel augmented by a density-dependent, zero-range pairing term of mixed type \cite{Dobaczewski2002} were chosen: SkM* \cite{Bartel1982}, SLy4 \cite{Chabanat1998}, SV-min \cite{Klupfel2009}, UNEDF0 \cite{Kortelainen2010}, UNEDF1 \cite{Kortelainen2012}, UNEDF2 \cite{Kortelainen2014}, and UNEDF1$^{\textnormal{SO}}$ \cite{Shi2014}. The functionals SkM* (developed with a focus on surface energy to properly account for the fission barrier of $^{240}$Pu using a semiclassical method) and SLy4 (developed with an emphasis on neutron-rich nuclei) are included for their value as traditional Skyrme EDFs, and serve as a benchmark against the performance of the newer parameterizations.  The EDF SV-min was parameterized with the binding energies, charge and diffraction radii, and surface thicknesses of semimagic nuclei. The UNEDF0 parameterization was optimized to the binding energies, charge radii, and odd-even binding energy differences of spherical and deformed nuclei. The EDF UNEDF1 was developed for fission studies and extended the data set of UNEDF0 with the inclusion of new masses and the excitation energies of fission isomers. The functional UNEDF2 considers the tensor terms ignored in the previous UNEDF parametrizations; it was developed for studies of shell structure and extended the data set of UNEDF1 with single-particle energy splittings of doubly-magic nuclei.  Finally, UNEDF1$^{\textnormal{SO}}$ is an EDF locally optimized in the transuranic region with the spin-orbit and pairing parameters fine-tuned to achieve a better agreement with both the excitation spectra and odd-even mass differences in $^{249}$Bk and $^{251}$Cf (with all of the other parameters being identical to UNEDF1).  The selection of several EDFs, based on different optimization methodologies, allows for an estimation of systematic errors.

The procedure we used to perform our calculations was identical to that in our previous work on nuclear drip lines \cite{Erler2012b}. For a given nucleus, we solved the Hartree-Fock-Bogoliubov (HFB) equations of nuclear DFT \cite{Ring1980} to find its ground-state binding energy and other global nuclear properties.  Given the impact of shape deformation on nuclear binding energy, it was necessary to solve the HFB equations for several different nuclear configurations; the nuclear deformation (and other global nuclear properties) corresponding to the minimum binding energy were then recorded for each nucleus and used in subsequent calculations. Since there were thousands of calculations to make for many different nuclei, we utilized high-performance computing to expedite the process.

As the focus of our work was on superheavy systems, we limited ourselves to nuclei with proton numbers $98 \leq Z \leq 120$. Also, we limited ourselves to nuclei with even numbers of protons and neutrons to avoid the complexities associated with odd-$A$ and odd-odd systems \cite{Afanasjev2015, Bonneau2007, Schunck2010}. To carry out our calculations, we used the DFT code HFBTHOv300 \cite{Navarro2017}, which solves the HFB equations through direct diagonalization in the deformed harmonic oscillator basis. We included constraints on the quadrupole deformation $\beta_2$ to account for prolate, oblate, and spherical deformations.  To expedite calculations, we imposed axial and reflection symmetry. Though the presence of triaxial shapes in this region is well established \cite{Cwiok2005}, their impact on the ground-state binding energy is predicted to be small \cite{Moller2008}, so this is a reasonable approximation. To approximately restore particle number symmetry broken in the HFB method, we used the Lipkin-Nogami procedure outlined in Ref.~\cite{Stoitsov2007}.

While for several EDFs the mass tables computed in Ref.~\cite{Erler2012b} have been stored in the database MassExplorer \cite{massexplorer}, in order to be sure that the quadrupole deformations do not  suddenly jump to extreme values, we recomputed all the mass tables and updated MassExplorer. From the calculated binding energies we extracted $Q_{\alpha}$ values:
\begin{equation}
  Q_{\alpha} = 28.3\,{\textnormal{MeV}} - \textnormal{BE}(Z,N) + \textnormal{BE}(Z-2,N-2).
\end{equation}

To assess the quality of our $Q_{\alpha}$ values we took two approaches. The first was to directly analyze the results from all 7 EDFs individually and compare them to one another and available experimental values. The second approach, used to estimate systematic uncertainties, was to mix several models.

\section{Results}\label{results}
\begin{figure}[htb]
  \includegraphics[width=0.9\linewidth]{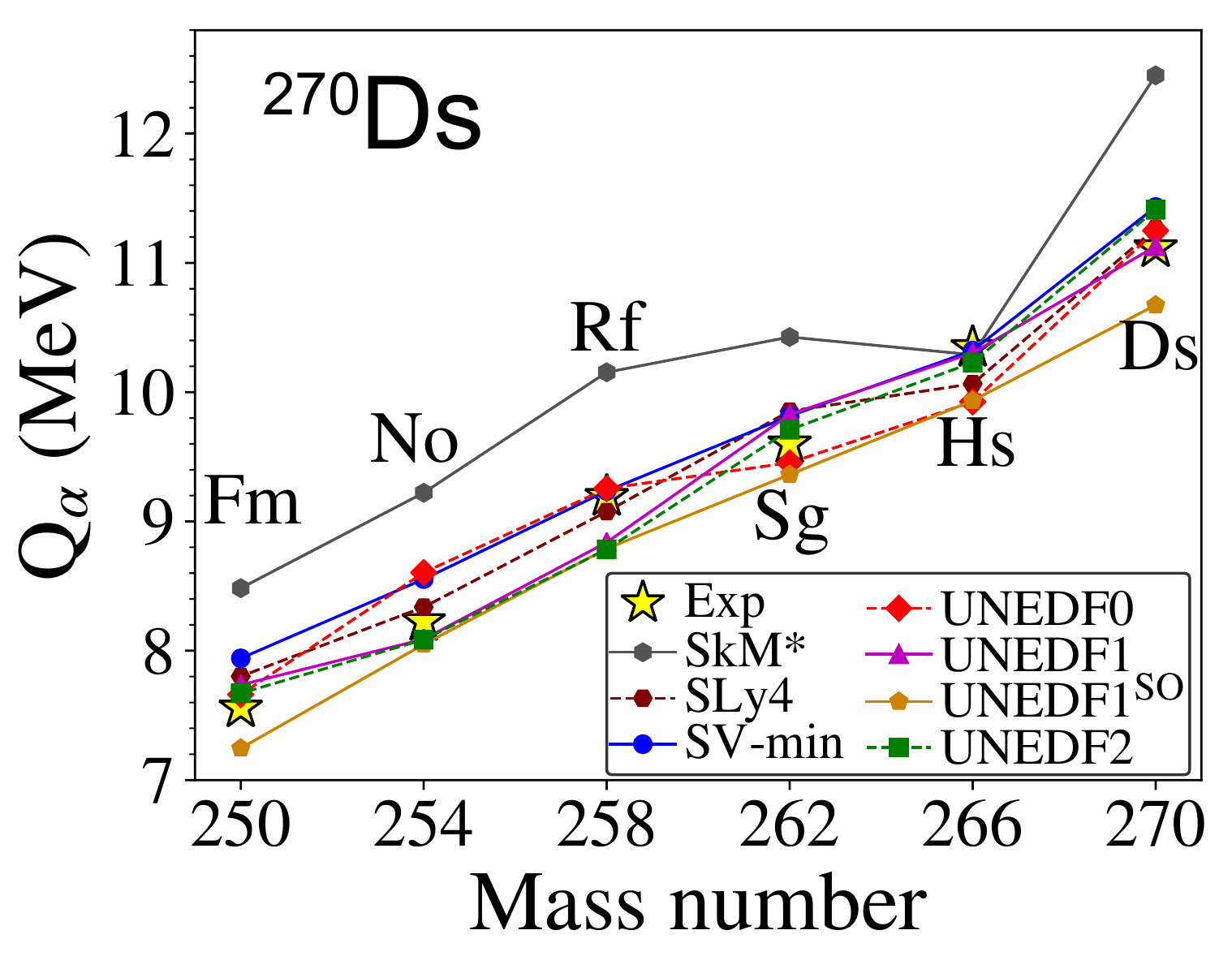}
  \caption{$Q_{\alpha}$ values for nuclei along the $\alpha$-decay chain of $^{270}$Ds computed with several Skyrme EDFs. Experimental values from Ref.~\cite{Wang2017} are represented as stars.}
  \label{alpha-decay_chain-270Ds}
\end{figure}

We begin by evaluating our calculations for the $\alpha$-decay chains of selected nuclei.  In Fig. \ref{alpha-decay_chain-270Ds} we compare our nuclear DFT results for the $Q_{\alpha}$ values of the $\alpha$-decay chain of $^{270}$Ds to experimental data; this nucleus was chosen for the availability of experimental data for every nucleus in its $\alpha$-decay chain. The first thing we observe here is the overall consistency of the Skyrme EDF results: with the exception of $^{266}$Hs for SkM*, every predicted $Q_{\alpha}$ value follows the pattern of experimental data and increases with increasing mass number. We also notice that, when excluding SkM*, the spread of calculated $Q_{\alpha}$ values is less than $1$ MeV, and each data point lies within this range ($^{266}$Hs being an exception, where it is $25$ keV above the closest result from SV-min). While it almost always overestimates the experimental data, the performance of SkM* here is not surprising given the improvement of later EDFs.

\begin{figure}[b]
  \includegraphics[width=0.9\linewidth]{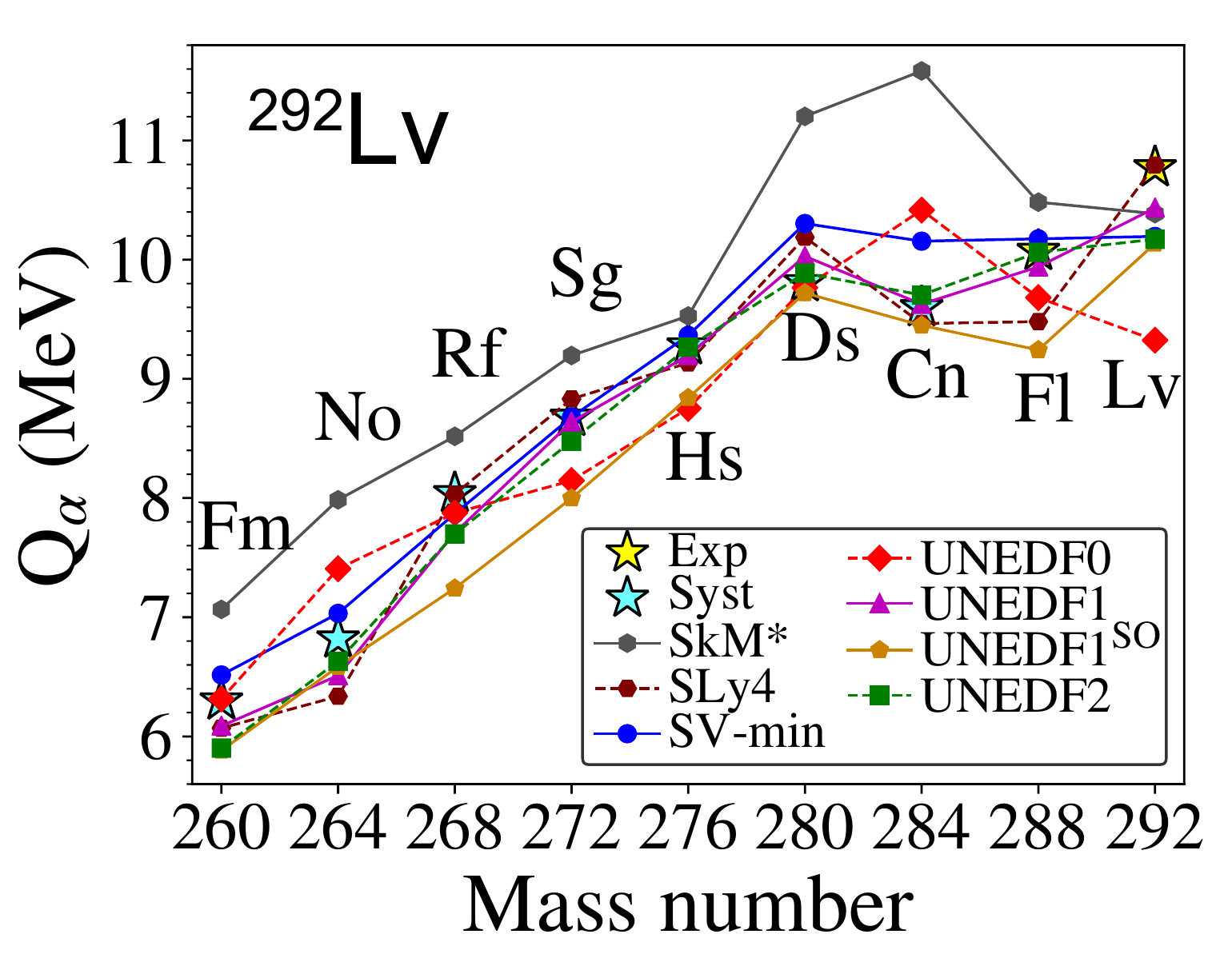}
  \caption{Similar to Fig. \ref{alpha-decay_chain-270Ds} but for $^{292}$Lv. Recommended values extrapolated from systematic trends from Ref.~\cite{Wang2017} (Syst) are also represented as stars.}
  \label{alpha-decay_chain-292Lv}
\end{figure}

In Fig. \ref{alpha-decay_chain-292Lv} we show our $Q_{\alpha}$ results for the $\alpha$-decay chain of $^{292}$Lv, where the large number of values extrapolated from systematic trends highlights the scarcity of experimental data in this region. Up to Ds, the pattern looks very similar to that of Fig. \ref{alpha-decay_chain-270Ds} for the given extrapolated values. However, at $^{284}$Cn there is a slight decrease in the extrapolated $Q_{\alpha}$ value, followed by an increase in the experimental $Q_{\alpha}$ value at $^{288}$Fl.  This is due to an abrupt shape transition from prolate to oblate deformation near $N=174$ caused by the triaxial softness of this region \cite{Cwiok2005, Heenen2015}. As a result, the consistency of the EDF results suffers, most noticeably with UNEDF0, whose results decrease from Fl to Lv while the experimental data increases. The reduced impact of the shape transition on UNEDF1 and UNEDF2 appears to highlight the necessity of including fission isomer and single-particle energy data in the global EDF optimization.

\begin{figure}[b]
  \includegraphics[width=0.9\columnwidth]{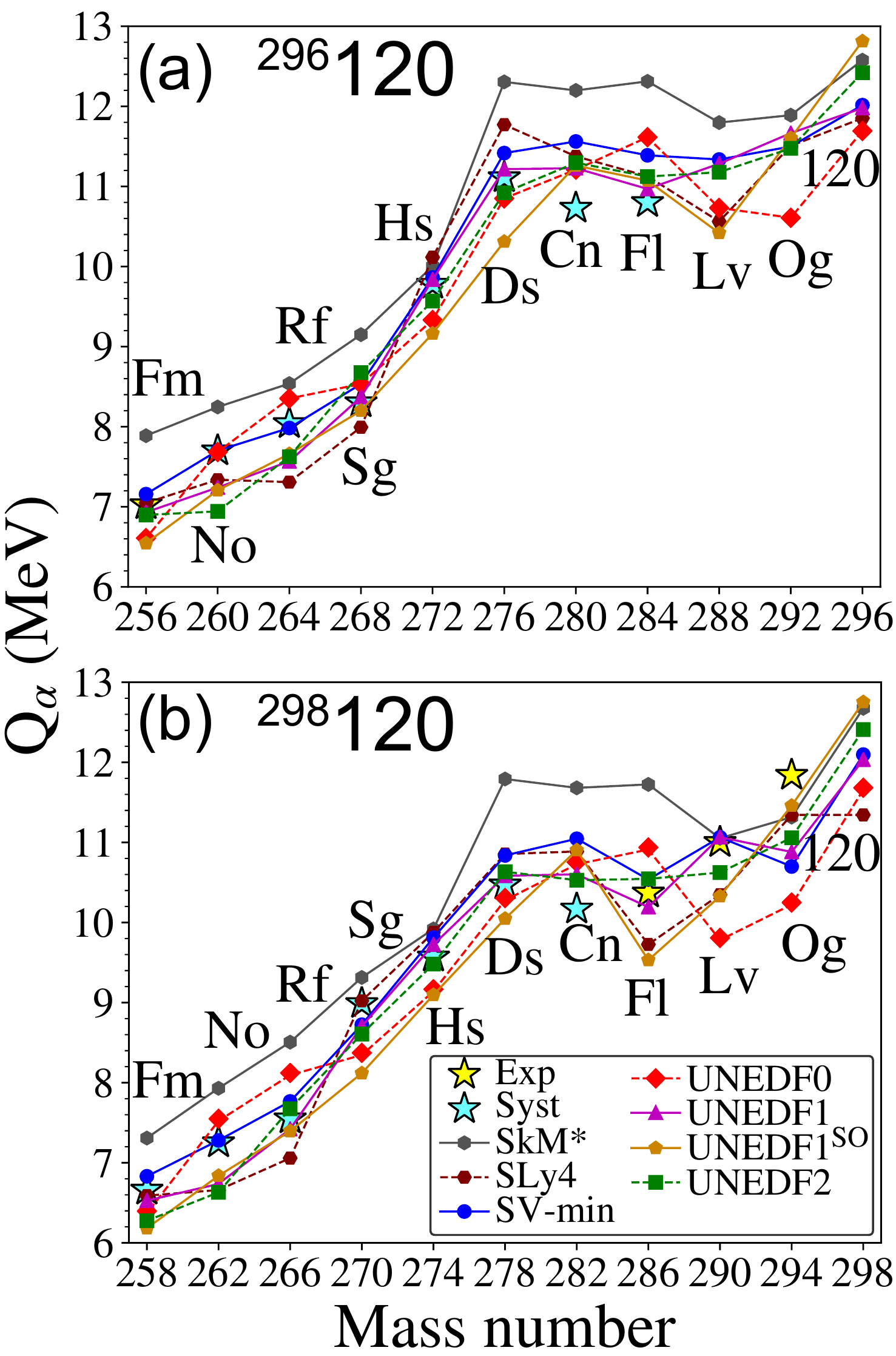}
  \caption{Similar to Figs.~\ref{alpha-decay_chain-270Ds} and \ref{alpha-decay_chain-292Lv} but for $^{296}$120 (a) and $^{298}$120 (b).}
  \label{alpha-decay_chain-296_298120}
\end{figure}

\begin{figure*}[htb]
  \includegraphics[width=\linewidth]{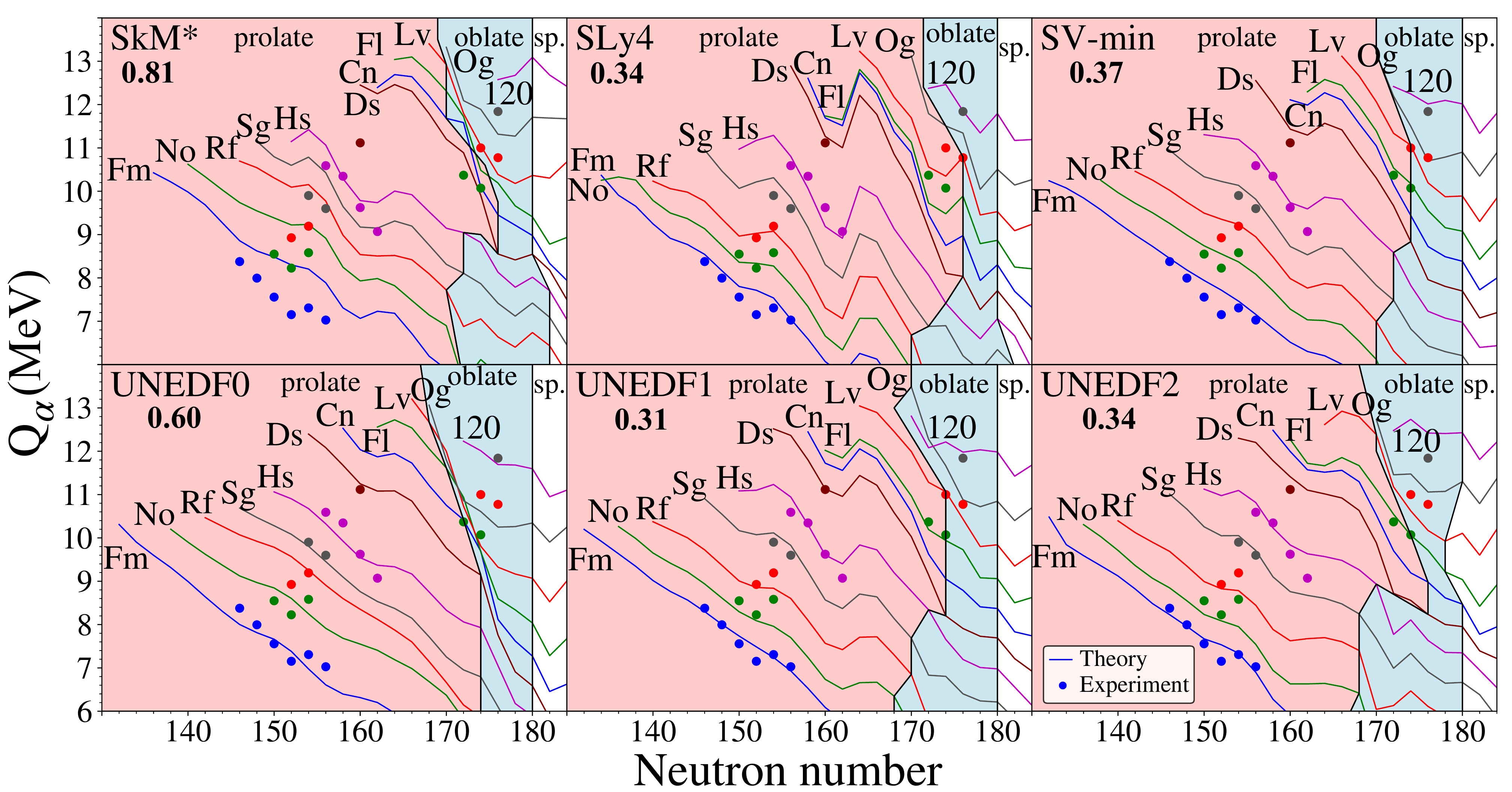}
  \caption{$Q_{\alpha}$ values for the isotopic chains of nuclei from Fm to $Z=120$ for each global Skyrme EDF used in this study. Areas of prolate, oblate, and spherical shapes are marked. Experimental data from Ref. \cite{Wang2017} are shown as circles and match the color of the corresponding line representing theoretical calculations (we note that no experimental data currently exist for the even-even isotopes of Cn). The root-mean-square (rms) deviation from experimental data (in MeV) is indicated for each model. For the local EDF UNEDF1$^{\textnormal{SO}}$ (not shown), the rms deviation was $0.46$ MeV.}
  \label{Isotopic_chain_plot_All}
\end{figure*}

We also want to extend our analysis to nuclei which have not yet been observed. For this, we show our results for the $Q_{\alpha}$ values of the $\alpha$-decay chains of $^{296}$120 and $^{298}$120 in Fig. \ref{alpha-decay_chain-296_298120}. Just like in Figs. \ref{alpha-decay_chain-270Ds} and \ref{alpha-decay_chain-292Lv}, the pattern from Fm to Ds is repeated for all of the EDFs and experiment. Upon reaching the prolate to oblate transition, we see a behavior similar to the results of Fig. \ref{alpha-decay_chain-292Lv}: for $^{296}$120, we notice the values obtained from systematic trends for $^{280}$Cn and $^{284}$Fl lie below the distribution of Skyrme EDF results, though the results of SLy4, UNEDF1, UNEDF1$^{\textnormal{SO}}$, and UNEDF2 for $^{284}$Fl are reasonably close. For $^{298}$120, we see again that the recommended value for $^{282}$Cn is below the range of calculated results, while the experimental data of $^{286}$Fl and $^{290}$Lv are within it (the lack of experimental data before $^{286}$Fl is due to the dominance of fission in $^{282}$Cn \cite{Brewer2018}, which prevents the measurement of $Q_{\alpha}$ values in this region). The most striking feature here is the comparison between the experimental data point and calculated results of $^{294}$Og, where every EDF underestimates the experimental value. The fact that UNEDF1$^{\textnormal{SO}}$ is closest to this value is likely due to its prediction of the shape transition from oblate deformation to spherical shape starting near $N=178$, and could be an indication of the importance of the spin-orbit force for this nucleus. Again, the significant underestimation of the experimental value by UNEDF0, in comparison with UNEDF1, UNEDF1$^{\textnormal{SO}}$, and UNEDF2, illustrates the need to incorporate more data into future EDFs. We also notice a trend in each calculated $Q_{\alpha}$ value to increase for both $^{296}$120 and $^{298}$120; given the pattern seen in experimental data from Fl to Og, this behavior is promising.


Figure~\ref{Isotopic_chain_plot_All} shows the analysis of $Q_{\alpha}$ values along the isotopic chains from Fm to $Z=120$. The borders between regions of prolate and oblate deformations and spherical shapes are marked. (See Ref.~\cite{Heenen2015} for discussion of deformation predictions in other models.) The irregularity seen around $N=164$, particularly well pronounced for SLy4, SV-min, and UNEDF1, is due to a prolate-deformed neutron subshell closure \cite{Cwiok1983, Moller1994}.
Once again, we observe an overall consistency for each EDF, with similar patterns emerging in the theoretical calculations for each isotopic chain, even in the regions where shape transitions occur. The proximity of our theoretical results to the experimental values, expressed through root-mean-square (rms) deviations $\delta(Q_{\alpha})$, is quite reasonable. As discussed earlier, the largest deviations from experiment are obtained for the heaviest elements Lv and Og, which are predicted to lie in the region of prolate-to-oblate shape transition. When inspecting the individual rms deviations, the best performer is UNEDF1 with $\delta(Q_{\alpha})=0.31$\,MeV, while the earliest Skyrme EDF SkM* yields $\delta(Q_{\alpha})=0.81$\,MeV. In general, the range of rms deviations here is consistent with that of other DFT models \cite{Heenen2015}. For instance, if one considers the relativistic EDFs of Ref.~\cite{Agbemava2015}, the $\delta(Q_{\alpha})$ values range from 0.32\,MeV for PC-PK1 to 0.68\,MeV for NL3*.

\begin{figure}[htb]
  \centering
  \includegraphics[width=0.9\linewidth]{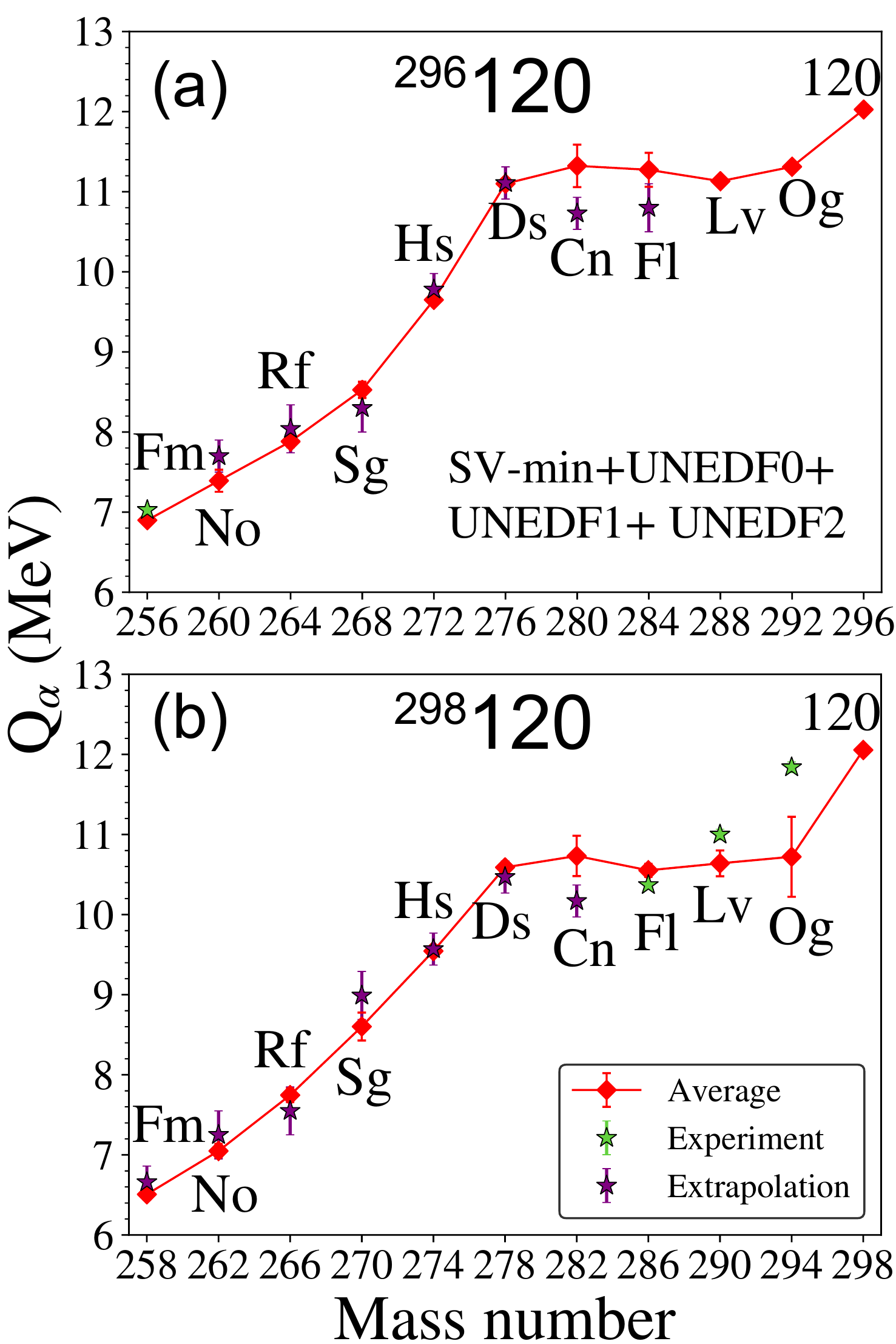}
  \caption{Model averaged $Q_{\alpha}$ values for the nuclei along the $\alpha$-decay chains of $^{296}$120 (a) and $^{298}$120 (b) calculated with the three UNEDF models and SV-min. Uniform model weights were assumed. Error bars on the theoretical predictions represent standard deviations. Experimental and recommended values from Ref.~\cite{Wang2017} are shown as stars with corresponding error bars.}
  \label{model_mixed_296298120}
\end{figure}

\begin{figure}[htb]
  \centering
  \includegraphics[width=0.9\linewidth]{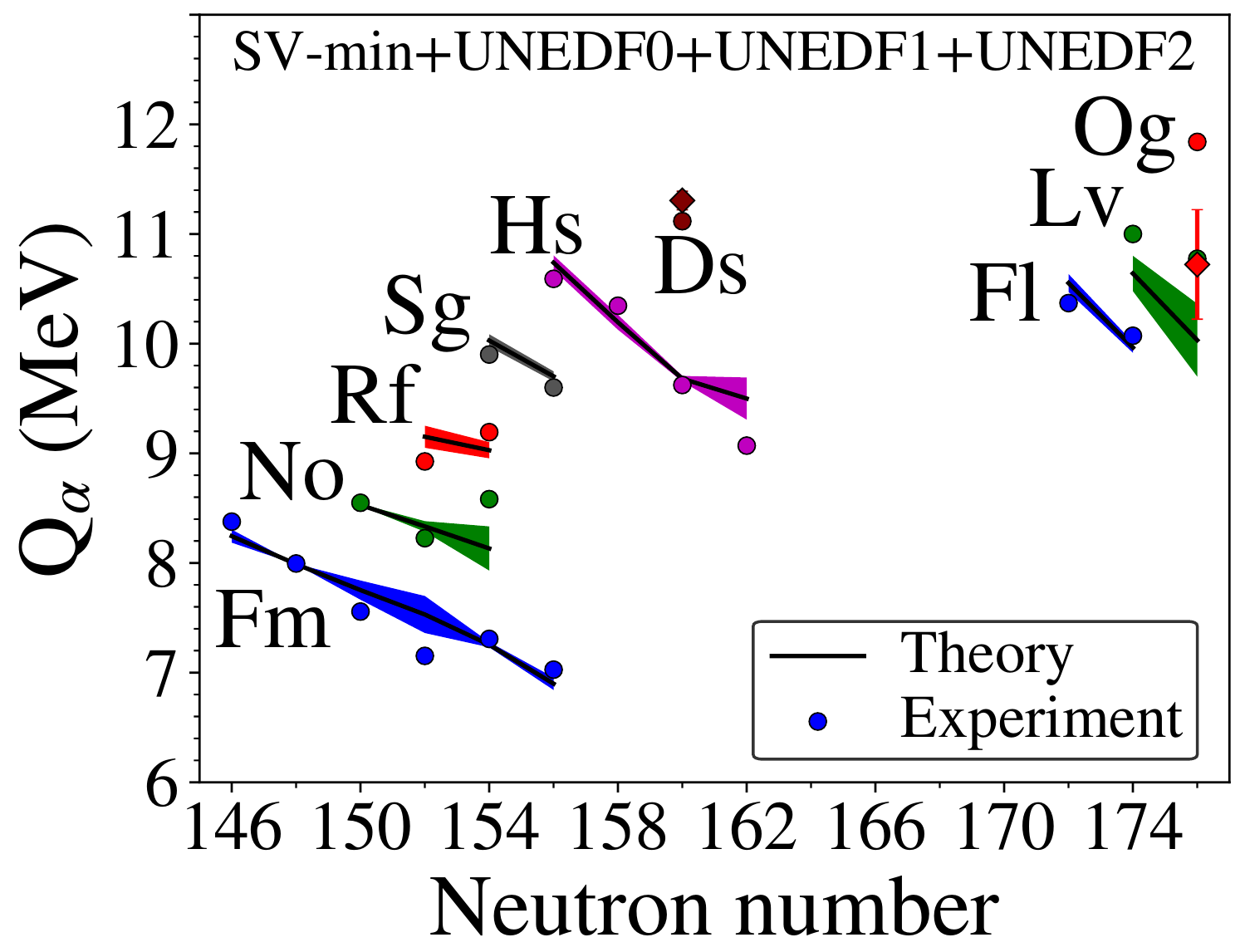}  
  \caption{Model-averaged $Q_{\alpha}$ values for the isotopic chains of even-even nuclei from Fm to $Z=120$ (excluding Cn) calculated with the three UNEDF models and SV-min. Uniform model weights were assumed. Error bars and error bands represent standard deviations. Experimental data from Ref.~\cite{Wang2017} are shown as circles and match the color of the line representing theoretical predictions.
  The rms deviation from experimental data for the model-averaged results is 0.35\,MeV.
  }
  \label{model_mixed_isotopic_chain}
\end{figure}

Assessing the uncertainty of a prediction made by a model in regions where experimental data are unavailable is a central issue in modern nuclear theory. So far we have estimated our uncertainty through the use of many different EDFs and compared their individual performances to experiment. Following Ref.~\cite{Erler2012b}, we now calculate the uniform average of the results of several models along with the corresponding standard deviations to determine the systematic uncertainty. For this we have chosen SV-min, UNEDF0, UNEDF1, and UNEDF2, as they are the most recently developed global EDFs used in this study. While this procedure may seem na{\"i}ve, without additional information or costly statistical calculations, the choice of uniform weights is essentially optimal \cite{Bernardo1994}. Also, by giving each model equal weight within the average, we can gain an idea of how more sophisticated model mixing may perform.

In Fig. \ref{model_mixed_296298120} we show the model averaged results for the $\alpha$-decay chains of $^{296}$120 and $^{298}$120. For both chains, between Fm and Ds, we see excellent agreement between calculated and experimental/recommended values. However, from Cn and beyond, the effects of the shape transition spoil this agreement. As discussed earlier, the difference is particularly noticeable for $^{294}$Og in the $^{298}$120 $\alpha$-decay chain; this is likely due to the low $Q_{\alpha}$ value predicted by UNEDF0 (see Fig. \ref{alpha-decay_chain-296_298120}). However, even when excluding UNEDF0 from the average (and including UNEDF1$^{\textnormal{SO}}$, whose $Q_{\alpha}$ value is much larger), the discrepancy remains substantial. In Fig. \ref{model_mixed_isotopic_chain}, we show the model-averaged results for the $Q_{\alpha}$ values of isotopic chains for which experimental data exist. From Fm to Fl, the proximity to experimental data is quite good, and the error bands are relatively small.  However, for Lv and Og, we see a similar behavior as in Fig.~\ref{model_mixed_296298120}.

\section{Conclusion}\label{conclusions}

In this paper we studied $Q_{\alpha}$ values for even-even superheavy nuclei from Fm to $Z=120$ within the framework of nuclear DFT with several different Skyrme EDFs. In order to estimate systematic errors, we analyzed theoretical predictions for $\alpha$-decay chains by comparing individual models, and also through model averaging. In the region of well deformed superheavy nuclei, the theoretical predictions are robust, with each EDF giving relatively consistent results. This robustness is somewhat reduced for shape-transitional nuclei. In general, the observed agreement with experimental data is quite reasonable.

The behavior of individual functionals, particularly from the UNEDF family, also proved enlightening. Among the models used, the best performer is UNEDF1 with an rms deviation of $\delta(Q_{\alpha})=0.31$\,MeV. The improvement in the results of UNEDF1 and UNEDF2 over UNEDF0 in the region of shape transition indicates the significance of data on fission isomers and one-quasiparticle states in the EDF optimization. We also analyzed the performance of the functional UNEDF1$^{\textnormal{SO}}$ that was locally optimized to the transuranic isotopes of Bk and Cf. Given its fine-tuning, it is interesting to see that its performance for $Q_{\alpha}$ values is similar, or slightly worse, as compared to the other UNEDF parametrizations.

In general, the method of nuclide identification through $Q_{\alpha}$ values is not expected to work well in the region of deformed-to-spherical shape transition. In this context, theory will benefit greatly from both progress in developing new spectroscopic-quality global EDFs and more sophisticated statistical techniques of uncertainty quantification.  Experimentally, work on identifying new superheavy nuclei from the upper superheavy (hot fusion) region, without the use of $Q_{\alpha}$ values, is already underway \cite{Gates2015, Gates2016, Gates2018}.

As the search continues for elements beyond Og \cite{Dullmann2016, Hessberger2017, Hofmann2016, Roberto2018}, the accurate calculation of $Q_{\alpha}$ values will prove more and more beneficial. The performance of our model mixing results in assessing and reducing uncertainty seems promising. Also, further improvements of predictability are expected through model mixing techniques of a more sophisticated type that utilize Bayesian model averaging \cite{Bernardo1994, Hoeting1999, Wasserman2000}, where the simple average is reweighted using model posterior probabilities computed by integrating the respective likelihoods over the parameter space. In the near future, however, to make more reliable extrapolations of $Q_{\alpha}$ values, we intend to use Bayesian machine learning techniques as described in the recent Ref.~ \cite{Neufcourt2018}.

\section{Acknowledgments}
Discussions with Nicolas Schunck are appreciated. The computational facility at Lawrence Livermore National Laboratory housing Quartz was instrumental in this work. This work was supported by the U.S. Department of Energy under Award Numbers DOE-DE-NA0002847 (NNSA, the Stewardship Science Academic Alliances program), DE-SC0013365 (Office of Science), and DE-SC0018083 (Office of Science, NUCLEI SciDAC-4 collaboration).

\bibliographystyle{apsrev4-1}
\bibliography{Alpha-Decay_references-corrected}

\begin{thebibliography}{88}%
\makeatletter
\providecommand \@ifxundefined [1]{%
 \@ifx{#1\undefined}
}%
\providecommand \@ifnum [1]{%
 \ifnum #1\expandafter \@firstoftwo
 \else \expandafter \@secondoftwo
 \fi
}%
\providecommand \@ifx [1]{%
 \ifx #1\expandafter \@firstoftwo
 \else \expandafter \@secondoftwo
 \fi
}%
\providecommand \natexlab [1]{#1}%
\providecommand \enquote  [1]{``#1''}%
\providecommand \bibnamefont  [1]{#1}%
\providecommand \bibfnamefont [1]{#1}%
\providecommand \citenamefont [1]{#1}%
\providecommand \href@noop [0]{\@secondoftwo}%
\providecommand \href [0]{\begingroup \@sanitize@url \@href}%
\providecommand \@href[1]{\@@startlink{#1}\@@href}%
\providecommand \@@href[1]{\endgroup#1\@@endlink}%
\providecommand \@sanitize@url [0]{\catcode `\\12\catcode `\$12\catcode
  `\&12\catcode `\#12\catcode `\^12\catcode `\_12\catcode `\%12\relax}%
\providecommand \@@startlink[1]{}%
\providecommand \@@endlink[0]{}%
\providecommand \url  [0]{\begingroup\@sanitize@url \@url }%
\providecommand \@url [1]{\endgroup\@href {#1}{\urlprefix }}%
\providecommand \urlprefix  [0]{URL }%
\providecommand \Eprint [0]{\href }%
\providecommand \doibase [0]{http://dx.doi.org/}%
\providecommand \selectlanguage [0]{\@gobble}%
\providecommand \bibinfo  [0]{\@secondoftwo}%
\providecommand \bibfield  [0]{\@secondoftwo}%
\providecommand \translation [1]{[#1]}%
\providecommand \BibitemOpen [0]{}%
\providecommand \bibitemStop [0]{}%
\providecommand \bibitemNoStop [0]{.\EOS\space}%
\providecommand \EOS [0]{\spacefactor3000\relax}%
\providecommand \BibitemShut  [1]{\csname bibitem#1\endcsname}%
\let\auto@bib@innerbib\@empty
\bibitem [{\citenamefont {Hoffman}\ \emph {et~al.}(2000)\citenamefont
  {Hoffman}, \citenamefont {Ghiorso},\ and\ \citenamefont
  {Seaborg}}]{Hoffman2000b}%
  \BibitemOpen
  \bibfield  {author} {\bibinfo {author} {\bibfnamefont {D.~C.}\ \bibnamefont
  {Hoffman}}, \bibinfo {author} {\bibfnamefont {A.}~\bibnamefont {Ghiorso}}, \
  and\ \bibinfo {author} {\bibfnamefont {G.~T.}\ \bibnamefont {Seaborg}},\
  }\href@noop {} {\emph {\bibinfo {title} {The Transuranium People: The Inside
  Story}}}\ (\bibinfo  {publisher} {Imperial College Press},\ \bibinfo {year}
  {2000})\BibitemShut {NoStop}%
\bibitem [{\citenamefont {Seaborg}\ and\ \citenamefont
  {Loveland}(1990)}]{Seaborg1990}%
  \BibitemOpen
  \bibfield  {author} {\bibinfo {author} {\bibfnamefont {G.~T.}\ \bibnamefont
  {Seaborg}}\ and\ \bibinfo {author} {\bibfnamefont {W.~D.}\ \bibnamefont
  {Loveland}},\ }\href@noop {} {\emph {\bibinfo {title} {The Elements Beyond
  Uranium}}}\ (\bibinfo  {publisher} {Wiley-Interscience},\ \bibinfo {year}
  {1990})\BibitemShut {NoStop}%
\bibitem [{\citenamefont {Giuliani}\ \emph {et~al.}(2018)\citenamefont
  {Giuliani}, \citenamefont {Matheson}, \citenamefont {Nazarewicz},
  \citenamefont {Olsen}, \citenamefont {Reinhard}, \citenamefont {Sadhukhan},
  \citenamefont {Schuetrumpf}, \citenamefont {Schunck},\ and\ \citenamefont
  {Schwerdtfeger}}]{Giuliani2018}%
  \BibitemOpen
  \bibfield  {author} {\bibinfo {author} {\bibfnamefont {S.}~\bibnamefont
  {Giuliani}}, \bibinfo {author} {\bibfnamefont {Z.}~\bibnamefont {Matheson}},
  \bibinfo {author} {\bibfnamefont {W.}~\bibnamefont {Nazarewicz}}, \bibinfo
  {author} {\bibfnamefont {E.}~\bibnamefont {Olsen}}, \bibinfo {author}
  {\bibfnamefont {P.-G.}\ \bibnamefont {Reinhard}}, \bibinfo {author}
  {\bibfnamefont {J.}~\bibnamefont {Sadhukhan}}, \bibinfo {author}
  {\bibfnamefont {B.}~\bibnamefont {Schuetrumpf}}, \bibinfo {author}
  {\bibfnamefont {N.}~\bibnamefont {Schunck}}, \ and\ \bibinfo {author}
  {\bibfnamefont {P.}~\bibnamefont {Schwerdtfeger}},\ }\href@noop {} {\bibfield
   {journal} {\bibinfo  {journal} {Rev. Mod. Phys.}\ } (\bibinfo {year}
  {2018})}\BibitemShut {NoStop}%
\bibitem [{\citenamefont {Nazarewicz}(2018)}]{Nazarewicz2018}%
  \BibitemOpen
  \bibfield  {author} {\bibinfo {author} {\bibfnamefont {W.}~\bibnamefont
  {Nazarewicz}},\ }\href {\doibase 10.1038/s41567-018-0163-3} {\bibfield
  {journal} {\bibinfo  {journal} {Nature Physics}\ }\textbf {\bibinfo {volume}
  {14}},\ \bibinfo {pages} {537} (\bibinfo {year} {2018})}\BibitemShut
  {NoStop}%
\bibitem [{\citenamefont {Viola}\ and\ \citenamefont
  {Seaborg}(1966)}]{Viola1966}%
  \BibitemOpen
  \bibfield  {author} {\bibinfo {author} {\bibfnamefont {V.}~\bibnamefont
  {Viola}}\ and\ \bibinfo {author} {\bibfnamefont {G.}~\bibnamefont
  {Seaborg}},\ }\href {\doibase https://doi.org/10.1016/0022-1902(66)80412-8}
  {\bibfield  {journal} {\bibinfo  {journal} {J. Inorg. Nucl. Chem.}\ }\textbf
  {\bibinfo {volume} {28}},\ \bibinfo {pages} {741 } (\bibinfo {year}
  {1966})}\BibitemShut {NoStop}%
\bibitem [{\citenamefont {Myers}\ and\ \citenamefont
  {Swiatecki}(1966)}]{Myers1966}%
  \BibitemOpen
  \bibfield  {author} {\bibinfo {author} {\bibfnamefont {W.~D.}\ \bibnamefont
  {Myers}}\ and\ \bibinfo {author} {\bibfnamefont {W.~J.}\ \bibnamefont
  {Swiatecki}},\ }\href {\doibase https://doi.org/10.1016/0029-5582(66)90639-0}
  {\bibfield  {journal} {\bibinfo  {journal} {Nucl. Phys.}\ }\textbf {\bibinfo
  {volume} {81}},\ \bibinfo {pages} {1 } (\bibinfo {year} {1966})}\BibitemShut
  {NoStop}%
\bibitem [{\citenamefont {Nilsson}\ \emph {et~al.}(1969)\citenamefont
  {Nilsson}, \citenamefont {Tsang}, \citenamefont {Sobiczewski}, \citenamefont
  {Szyma{\'n}ski}, \citenamefont {Wycech}, \citenamefont {Gustafson},
  \citenamefont {Lamm}, \citenamefont {M{\"o}ller},\ and\ \citenamefont
  {Nilsson}}]{Nilsson1969}%
  \BibitemOpen
  \bibfield  {author} {\bibinfo {author} {\bibfnamefont {S.~G.}\ \bibnamefont
  {Nilsson}}, \bibinfo {author} {\bibfnamefont {C.~F.}\ \bibnamefont {Tsang}},
  \bibinfo {author} {\bibfnamefont {A.}~\bibnamefont {Sobiczewski}}, \bibinfo
  {author} {\bibfnamefont {Z.}~\bibnamefont {Szyma{\'n}ski}}, \bibinfo {author}
  {\bibfnamefont {S.}~\bibnamefont {Wycech}}, \bibinfo {author} {\bibfnamefont
  {C.}~\bibnamefont {Gustafson}}, \bibinfo {author} {\bibfnamefont {I.-L.}\
  \bibnamefont {Lamm}}, \bibinfo {author} {\bibfnamefont {P.}~\bibnamefont
  {M{\"o}ller}}, \ and\ \bibinfo {author} {\bibfnamefont {B.}~\bibnamefont
  {Nilsson}},\ }\href {\doibase https://doi.org/10.1016/0375-9474(69)90809-4}
  {\bibfield  {journal} {\bibinfo  {journal} {Nucl. Phys. A}\ }\textbf
  {\bibinfo {volume} {131}},\ \bibinfo {pages} {1 } (\bibinfo {year}
  {1969})}\BibitemShut {NoStop}%
\bibitem [{\citenamefont {Sobiczewski}\ \emph {et~al.}(1966)\citenamefont
  {Sobiczewski}, \citenamefont {Gareev},\ and\ \citenamefont
  {Kalinkin}}]{Sobiczewski1966}%
  \BibitemOpen
  \bibfield  {author} {\bibinfo {author} {\bibfnamefont {A.}~\bibnamefont
  {Sobiczewski}}, \bibinfo {author} {\bibfnamefont {F.}~\bibnamefont {Gareev}},
  \ and\ \bibinfo {author} {\bibfnamefont {B.}~\bibnamefont {Kalinkin}},\
  }\href {\doibase https://doi.org/10.1016/0031-9163(66)91243-1} {\bibfield
  {journal} {\bibinfo  {journal} {Phys. Lett.}\ }\textbf {\bibinfo {volume}
  {22}},\ \bibinfo {pages} {500 } (\bibinfo {year} {1966})}\BibitemShut
  {NoStop}%
\bibitem [{\citenamefont {Afanasjev}\ \emph {et~al.}(2003)\citenamefont
  {Afanasjev}, \citenamefont {Khoo}, \citenamefont {Frauendorf}, \citenamefont
  {Lalazissis},\ and\ \citenamefont {Ahmad}}]{Afanasjev2003}%
  \BibitemOpen
  \bibfield  {author} {\bibinfo {author} {\bibfnamefont {A.~V.}\ \bibnamefont
  {Afanasjev}}, \bibinfo {author} {\bibfnamefont {T.~L.}\ \bibnamefont {Khoo}},
  \bibinfo {author} {\bibfnamefont {S.}~\bibnamefont {Frauendorf}}, \bibinfo
  {author} {\bibfnamefont {G.~A.}\ \bibnamefont {Lalazissis}}, \ and\ \bibinfo
  {author} {\bibfnamefont {I.}~\bibnamefont {Ahmad}},\ }\href {\doibase
  10.1103/PhysRevC.67.024309} {\bibfield  {journal} {\bibinfo  {journal} {Phys.
  Rev. C}\ }\textbf {\bibinfo {volume} {67}},\ \bibinfo {pages} {024309}
  (\bibinfo {year} {2003})}\BibitemShut {NoStop}%
\bibitem [{\citenamefont {Afanasjev}(2006)}]{Afanasjev2006}%
  \BibitemOpen
  \bibfield  {author} {\bibinfo {author} {\bibfnamefont {A.~V.}\ \bibnamefont
  {Afanasjev}},\ }\href {http://stacks.iop.org/1402-4896/2006/i=T125/a=014}
  {\bibfield  {journal} {\bibinfo  {journal} {Phys. Scr.}\ }\textbf {\bibinfo
  {volume} {2006}},\ \bibinfo {pages} {62} (\bibinfo {year}
  {2006})}\BibitemShut {NoStop}%
\bibitem [{\citenamefont {Bender}\ \emph {et~al.}(1999)\citenamefont {Bender},
  \citenamefont {Rutz}, \citenamefont {Reinhard}, \citenamefont {Maruhn},\ and\
  \citenamefont {Greiner}}]{Bender1999}%
  \BibitemOpen
  \bibfield  {author} {\bibinfo {author} {\bibfnamefont {M.}~\bibnamefont
  {Bender}}, \bibinfo {author} {\bibfnamefont {K.}~\bibnamefont {Rutz}},
  \bibinfo {author} {\bibfnamefont {P.-G.}\ \bibnamefont {Reinhard}}, \bibinfo
  {author} {\bibfnamefont {J.~A.}\ \bibnamefont {Maruhn}}, \ and\ \bibinfo
  {author} {\bibfnamefont {W.}~\bibnamefont {Greiner}},\ }\href {\doibase
  10.1103/PhysRevC.60.034304} {\bibfield  {journal} {\bibinfo  {journal} {Phys.
  Rev. C}\ }\textbf {\bibinfo {volume} {60}},\ \bibinfo {pages} {034304}
  (\bibinfo {year} {1999})}\BibitemShut {NoStop}%
\bibitem [{\citenamefont {{\'C}wiok}\ \emph {et~al.}(1996)\citenamefont
  {{\'C}wiok}, \citenamefont {Dobaczewski}, \citenamefont {Heenen},
  \citenamefont {Magierski},\ and\ \citenamefont {Nazarewicz}}]{Cwiok1996}%
  \BibitemOpen
  \bibfield  {author} {\bibinfo {author} {\bibfnamefont {S.}~\bibnamefont
  {{\'C}wiok}}, \bibinfo {author} {\bibfnamefont {J.}~\bibnamefont
  {Dobaczewski}}, \bibinfo {author} {\bibfnamefont {P.-H.}\ \bibnamefont
  {Heenen}}, \bibinfo {author} {\bibfnamefont {P.}~\bibnamefont {Magierski}}, \
  and\ \bibinfo {author} {\bibfnamefont {W.}~\bibnamefont {Nazarewicz}},\
  }\href {\doibase https://doi.org/10.1016/S0375-9474(96)00337-5} {\bibfield
  {journal} {\bibinfo  {journal} {Nucl. Phys. A}\ }\textbf {\bibinfo {volume}
  {611}},\ \bibinfo {pages} {211 } (\bibinfo {year} {1996})}\BibitemShut
  {NoStop}%
\bibitem [{\citenamefont {Kruppa}\ \emph {et~al.}(2000)\citenamefont {Kruppa},
  \citenamefont {Bender}, \citenamefont {Nazarewicz}, \citenamefont {Reinhard},
  \citenamefont {Vertse},\ and\ \citenamefont {\ifmmode~\acute{C}\else
  \'{C}\fi{}wiok}}]{Kruppa2000}%
  \BibitemOpen
  \bibfield  {author} {\bibinfo {author} {\bibfnamefont {A.~T.}\ \bibnamefont
  {Kruppa}}, \bibinfo {author} {\bibfnamefont {M.}~\bibnamefont {Bender}},
  \bibinfo {author} {\bibfnamefont {W.}~\bibnamefont {Nazarewicz}}, \bibinfo
  {author} {\bibfnamefont {P.-G.}\ \bibnamefont {Reinhard}}, \bibinfo {author}
  {\bibfnamefont {T.}~\bibnamefont {Vertse}}, \ and\ \bibinfo {author}
  {\bibfnamefont {S.}~\bibnamefont {\ifmmode~\acute{C}\else \'{C}\fi{}wiok}},\
  }\href {\doibase 10.1103/PhysRevC.61.034313} {\bibfield  {journal} {\bibinfo
  {journal} {Phys. Rev. C}\ }\textbf {\bibinfo {volume} {61}},\ \bibinfo
  {pages} {034313} (\bibinfo {year} {2000})}\BibitemShut {NoStop}%
\bibitem [{\citenamefont {Rutz}\ \emph {et~al.}(1997)\citenamefont {Rutz},
  \citenamefont {Bender}, \citenamefont {B\"urvenich}, \citenamefont
  {Schilling}, \citenamefont {Reinhard}, \citenamefont {Maruhn},\ and\
  \citenamefont {Greiner}}]{Rutz1997}%
  \BibitemOpen
  \bibfield  {author} {\bibinfo {author} {\bibfnamefont {K.}~\bibnamefont
  {Rutz}}, \bibinfo {author} {\bibfnamefont {M.}~\bibnamefont {Bender}},
  \bibinfo {author} {\bibfnamefont {T.}~\bibnamefont {B\"urvenich}}, \bibinfo
  {author} {\bibfnamefont {T.}~\bibnamefont {Schilling}}, \bibinfo {author}
  {\bibfnamefont {P.-G.}\ \bibnamefont {Reinhard}}, \bibinfo {author}
  {\bibfnamefont {J.~A.}\ \bibnamefont {Maruhn}}, \ and\ \bibinfo {author}
  {\bibfnamefont {W.}~\bibnamefont {Greiner}},\ }\href {\doibase
  10.1103/PhysRevC.56.238} {\bibfield  {journal} {\bibinfo  {journal} {Phys.
  Rev. C}\ }\textbf {\bibinfo {volume} {56}},\ \bibinfo {pages} {238} (\bibinfo
  {year} {1997})}\BibitemShut {NoStop}%
\bibitem [{\citenamefont {D\"{u}llmann}\ and\ \citenamefont
  {Block}(2018)}]{Dullmann2018}%
  \BibitemOpen
  \bibfield  {author} {\bibinfo {author} {\bibfnamefont {C.~E.}\ \bibnamefont
  {D\"{u}llmann}}\ and\ \bibinfo {author} {\bibfnamefont {M.}~\bibnamefont
  {Block}},\ }\href {\doibase 10.1038/scientificamerican0318-46} {\bibfield
  {journal} {\bibinfo  {journal} {Sci. Am.}\ }\textbf {\bibinfo {volume}
  {318}},\ \bibinfo {pages} {46} (\bibinfo {year} {2018})}\BibitemShut
  {NoStop}%
\bibitem [{\citenamefont {Agbemava}\ \emph {et~al.}(2015)\citenamefont
  {Agbemava}, \citenamefont {Afanasjev}, \citenamefont {Nakatsukasa},\ and\
  \citenamefont {Ring}}]{Agbemava2015}%
  \BibitemOpen
  \bibfield  {author} {\bibinfo {author} {\bibfnamefont {S.~E.}\ \bibnamefont
  {Agbemava}}, \bibinfo {author} {\bibfnamefont {A.~V.}\ \bibnamefont
  {Afanasjev}}, \bibinfo {author} {\bibfnamefont {T.}~\bibnamefont
  {Nakatsukasa}}, \ and\ \bibinfo {author} {\bibfnamefont {P.}~\bibnamefont
  {Ring}},\ }\href {\doibase 10.1103/PhysRevC.92.054310} {\bibfield  {journal}
  {\bibinfo  {journal} {Phys. Rev. C}\ }\textbf {\bibinfo {volume} {92}},\
  \bibinfo {pages} {054310} (\bibinfo {year} {2015})}\BibitemShut {NoStop}%
\bibitem [{\citenamefont {Bender}\ \emph {et~al.}(2001)\citenamefont {Bender},
  \citenamefont {Nazarewicz},\ and\ \citenamefont {Reinhard}}]{Bender2001}%
  \BibitemOpen
  \bibfield  {author} {\bibinfo {author} {\bibfnamefont {M.}~\bibnamefont
  {Bender}}, \bibinfo {author} {\bibfnamefont {W.}~\bibnamefont {Nazarewicz}},
  \ and\ \bibinfo {author} {\bibfnamefont {P.-G.}\ \bibnamefont {Reinhard}},\
  }\href {\doibase 0.1016/S0370-2693(01)00863-2} {\bibfield  {journal}
  {\bibinfo  {journal} {Phys. Lett. B}\ }\textbf {\bibinfo {volume} {515}},\
  \bibinfo {pages} {42 } (\bibinfo {year} {2001})}\BibitemShut {NoStop}%
\bibitem [{\citenamefont {Jerabek}\ \emph {et~al.}(2018)\citenamefont
  {Jerabek}, \citenamefont {Schuetrumpf}, \citenamefont {Schwerdtfeger},\ and\
  \citenamefont {Nazarewicz}}]{Jerabek2018}%
  \BibitemOpen
  \bibfield  {author} {\bibinfo {author} {\bibfnamefont {P.}~\bibnamefont
  {Jerabek}}, \bibinfo {author} {\bibfnamefont {B.}~\bibnamefont
  {Schuetrumpf}}, \bibinfo {author} {\bibfnamefont {P.}~\bibnamefont
  {Schwerdtfeger}}, \ and\ \bibinfo {author} {\bibfnamefont {W.}~\bibnamefont
  {Nazarewicz}},\ }\href {\doibase 10.1103/PhysRevLett.120.053001} {\bibfield
  {journal} {\bibinfo  {journal} {Phys. Rev. Lett.}\ }\textbf {\bibinfo
  {volume} {120}},\ \bibinfo {pages} {053001} (\bibinfo {year}
  {2018})}\BibitemShut {NoStop}%
\bibitem [{\citenamefont {Oganessian}\ \emph {et~al.}(1975)\citenamefont
  {Oganessian}, \citenamefont {Demin}, \citenamefont {Iljinov}, \citenamefont
  {Tretyakova}, \citenamefont {Pleve}, \citenamefont {Penionzhkevich},
  \citenamefont {Ivanov},\ and\ \citenamefont {Tretyakov}}]{Oganessian1975}%
  \BibitemOpen
  \bibfield  {author} {\bibinfo {author} {\bibfnamefont {Y.}~\bibnamefont
  {Oganessian}}, \bibinfo {author} {\bibfnamefont {A.}~\bibnamefont {Demin}},
  \bibinfo {author} {\bibfnamefont {A.}~\bibnamefont {Iljinov}}, \bibinfo
  {author} {\bibfnamefont {S.}~\bibnamefont {Tretyakova}}, \bibinfo {author}
  {\bibfnamefont {A.}~\bibnamefont {Pleve}}, \bibinfo {author} {\bibfnamefont
  {Y.}~\bibnamefont {Penionzhkevich}}, \bibinfo {author} {\bibfnamefont
  {M.}~\bibnamefont {Ivanov}}, \ and\ \bibinfo {author} {\bibfnamefont
  {Y.}~\bibnamefont {Tretyakov}},\ }\href {\doibase
  https://doi.org/10.1016/0375-9474(75)91140-9} {\bibfield  {journal} {\bibinfo
   {journal} {Nucl. Phys. A}\ }\textbf {\bibinfo {volume} {239}},\ \bibinfo
  {pages} {157 } (\bibinfo {year} {1975})}\BibitemShut {NoStop}%
\bibitem [{\citenamefont {Oganessian}(2007)}]{Oganessian2007}%
  \BibitemOpen
  \bibfield  {author} {\bibinfo {author} {\bibfnamefont {Y.}~\bibnamefont
  {Oganessian}},\ }\href {http://stacks.iop.org/0954-3899/34/i=4/a=R01}
  {\bibfield  {journal} {\bibinfo  {journal} {J. Phys. G}\ }\textbf {\bibinfo
  {volume} {34}},\ \bibinfo {pages} {R165} (\bibinfo {year}
  {2007})}\BibitemShut {NoStop}%
\bibitem [{\citenamefont {Barber}\ \emph {et~al.}(2011)\citenamefont {Barber},
  \citenamefont {Karol}, \citenamefont {Nakahara}, \citenamefont {E.Vardaci},\
  and\ \citenamefont {Vogt}}]{Barber2011}%
  \BibitemOpen
  \bibfield  {author} {\bibinfo {author} {\bibfnamefont {R.}~\bibnamefont
  {Barber}}, \bibinfo {author} {\bibfnamefont {P.~J.}\ \bibnamefont {Karol}},
  \bibinfo {author} {\bibfnamefont {H.}~\bibnamefont {Nakahara}}, \bibinfo
  {author} {\bibnamefont {E.Vardaci}}, \ and\ \bibinfo {author} {\bibfnamefont
  {E.~W.}\ \bibnamefont {Vogt}},\ }\href {\doibase 10.1351/PAC-REP-10-05-01}
  {\bibfield  {journal} {\bibinfo  {journal} {Pure Appl. Chem.}\ }\textbf
  {\bibinfo {volume} {83}},\ \bibinfo {pages} {1485} (\bibinfo {year}
  {2011})}\BibitemShut {NoStop}%
\bibitem [{\citenamefont {Karol}\ \emph
  {et~al.}(2016{\natexlab{a}})\citenamefont {Karol}, \citenamefont {Barber},
  \citenamefont {Sherrill}, \citenamefont {Emanuele},\ and\ \citenamefont
  {Toshimitsu}}]{Karol2016a}%
  \BibitemOpen
  \bibfield  {author} {\bibinfo {author} {\bibfnamefont {P.~J.}\ \bibnamefont
  {Karol}}, \bibinfo {author} {\bibfnamefont {R.~C.}\ \bibnamefont {Barber}},
  \bibinfo {author} {\bibfnamefont {B.~M.}\ \bibnamefont {Sherrill}}, \bibinfo
  {author} {\bibfnamefont {V.}~\bibnamefont {Emanuele}}, \ and\ \bibinfo
  {author} {\bibfnamefont {Y.}~\bibnamefont {Toshimitsu}},\ }\href {\doibase
  10.1515/pac-2015-0502} {\bibfield  {journal} {\bibinfo  {journal} {Pure Appl.
  Chem.}\ }\textbf {\bibinfo {volume} {88}},\ \bibinfo {pages} {139} (\bibinfo
  {year} {2016}{\natexlab{a}})}\BibitemShut {NoStop}%
\bibitem [{\citenamefont {Karol}\ \emph
  {et~al.}(2016{\natexlab{b}})\citenamefont {Karol}, \citenamefont {Barber},
  \citenamefont {Sherrill}, \citenamefont {Emanuele},\ and\ \citenamefont
  {Toshimitsu}}]{Karol2016b}%
  \BibitemOpen
  \bibfield  {author} {\bibinfo {author} {\bibfnamefont {P.~J.}\ \bibnamefont
  {Karol}}, \bibinfo {author} {\bibfnamefont {R.~C.}\ \bibnamefont {Barber}},
  \bibinfo {author} {\bibfnamefont {B.~M.}\ \bibnamefont {Sherrill}}, \bibinfo
  {author} {\bibfnamefont {V.}~\bibnamefont {Emanuele}}, \ and\ \bibinfo
  {author} {\bibfnamefont {Y.}~\bibnamefont {Toshimitsu}},\ }\href {\doibase
  10.1515/pac-2015-0501} {\bibfield  {journal} {\bibinfo  {journal} {Pure Appl.
  Chem.}\ }\textbf {\bibinfo {volume} {88}},\ \bibinfo {pages} {155} (\bibinfo
  {year} {2016}{\natexlab{b}})}\BibitemShut {NoStop}%
\bibitem [{\citenamefont {D\"{u}llmann}(2011)}]{Dullmann2011}%
  \BibitemOpen
  \bibfield  {author} {\bibinfo {author} {\bibfnamefont {C.~E.}\ \bibnamefont
  {D\"{u}llmann}},\ }\href
  {http://iopscience.iop.org/article/10.1088/1402-4896/aa53c1/pdf} {\bibfield
  {journal} {\bibinfo  {journal} {GSI Scientific Report 2011}\ }\textbf
  {\bibinfo {volume} {GSI Report 2012-1}},\ \bibinfo {pages} {206} (\bibinfo
  {year} {2011})}\BibitemShut {NoStop}%
\bibitem [{\citenamefont {D\"{u}llmann}(2017)}]{Dullmann2017}%
  \BibitemOpen
  \bibfield  {author} {\bibinfo {author} {\bibfnamefont {C.~E.}\ \bibnamefont
  {D\"{u}llmann}},\ }\href {\doibase 10.1051/epjconf/201716300015} {\bibfield
  {journal} {\bibinfo  {journal} {EPJ Web Conf.}\ }\textbf {\bibinfo {volume}
  {163}},\ \bibinfo {pages} {00015} (\bibinfo {year} {2017})}\BibitemShut
  {NoStop}%
\bibitem [{\citenamefont {Oganessian}\ \emph {et~al.}(2009)\citenamefont
  {Oganessian} \emph {et~al.}}]{Oganessian2009}%
  \BibitemOpen
  \bibfield  {author} {\bibinfo {author} {\bibfnamefont {Y.~T.}\ \bibnamefont
  {Oganessian}} \emph {et~al.},\ }\href {\doibase 10.1103/PhysRevC.79.024603}
  {\bibfield  {journal} {\bibinfo  {journal} {Phys. Rev. C}\ }\textbf {\bibinfo
  {volume} {79}},\ \bibinfo {pages} {024603} (\bibinfo {year}
  {2009})}\BibitemShut {NoStop}%
\bibitem [{\citenamefont {Hofmann}\ \emph {et~al.}(2004)\citenamefont
  {Hofmann}, \citenamefont {He{\ss}berger}, \citenamefont {Ackermann},
  \citenamefont {Antalic}, \citenamefont {Cagarda}, \citenamefont {Kindler},
  \citenamefont {Kuusiniemi}, \citenamefont {Leino}, \citenamefont {Lommel},
  \citenamefont {Malyshev}, \citenamefont {Mann}, \citenamefont {Mu¨nzenberg},
  \citenamefont {Popeko}, \citenamefont {S\'{}aro}, \citenamefont {Streicher},\
  and\ \citenamefont {Yeremin}}]{Hofmann2004}%
  \BibitemOpen
  \bibfield  {author} {\bibinfo {author} {\bibfnamefont {S.}~\bibnamefont
  {Hofmann}}, \bibinfo {author} {\bibfnamefont {F.}~\bibnamefont
  {He{\ss}berger}}, \bibinfo {author} {\bibfnamefont {D.}~\bibnamefont
  {Ackermann}}, \bibinfo {author} {\bibfnamefont {S.}~\bibnamefont {Antalic}},
  \bibinfo {author} {\bibfnamefont {P.}~\bibnamefont {Cagarda}}, \bibinfo
  {author} {\bibfnamefont {B.}~\bibnamefont {Kindler}}, \bibinfo {author}
  {\bibfnamefont {P.}~\bibnamefont {Kuusiniemi}}, \bibinfo {author}
  {\bibfnamefont {M.}~\bibnamefont {Leino}}, \bibinfo {author} {\bibfnamefont
  {B.}~\bibnamefont {Lommel}}, \bibinfo {author} {\bibfnamefont
  {O.}~\bibnamefont {Malyshev}}, \bibinfo {author} {\bibfnamefont
  {R.}~\bibnamefont {Mann}}, \bibinfo {author} {\bibfnamefont {G.}~\bibnamefont
  {Mu¨nzenberg}}, \bibinfo {author} {\bibfnamefont {A.}~\bibnamefont
  {Popeko}}, \bibinfo {author} {\bibfnamefont {S.}~\bibnamefont {S\'{}aro}},
  \bibinfo {author} {\bibfnamefont {B.}~\bibnamefont {Streicher}}, \ and\
  \bibinfo {author} {\bibfnamefont {A.}~\bibnamefont {Yeremin}},\ }\href
  {\doibase https://doi.org/10.1016/j.nuclphysa.2004.01.018} {\bibfield
  {journal} {\bibinfo  {journal} {Nucl. Phys. A}\ }\textbf {\bibinfo {volume}
  {734}},\ \bibinfo {pages} {93 } (\bibinfo {year} {2004})}\BibitemShut
  {NoStop}%
\bibitem [{\citenamefont {Oganessian}\ and\ \citenamefont
  {Utyonkov}(2015)}]{Oganessian2015}%
  \BibitemOpen
  \bibfield  {author} {\bibinfo {author} {\bibfnamefont {Y.~T.}\ \bibnamefont
  {Oganessian}}\ and\ \bibinfo {author} {\bibfnamefont {V.~K.}\ \bibnamefont
  {Utyonkov}},\ }\href {http://stacks.iop.org/0034-4885/78/i=3/a=036301}
  {\bibfield  {journal} {\bibinfo  {journal} {Rep. Prog. Phys.}\ }\textbf
  {\bibinfo {volume} {78}},\ \bibinfo {pages} {036301} (\bibinfo {year}
  {2015})}\BibitemShut {NoStop}%
\bibitem [{\citenamefont {Oganessian}\ \emph {et~al.}(2017)\citenamefont
  {Oganessian}, \citenamefont {Sobiczewski},\ and\ \citenamefont
  {Ter-Akopian}}]{Oganessian2017}%
  \BibitemOpen
  \bibfield  {author} {\bibinfo {author} {\bibfnamefont {Y.~T.}\ \bibnamefont
  {Oganessian}}, \bibinfo {author} {\bibfnamefont {A.}~\bibnamefont
  {Sobiczewski}}, \ and\ \bibinfo {author} {\bibfnamefont {G.~M.}\ \bibnamefont
  {Ter-Akopian}},\ }\href {http://stacks.iop.org/1402-4896/92/i=2/a=023003}
  {\bibfield  {journal} {\bibinfo  {journal} {Phys. Scr.}\ }\textbf {\bibinfo
  {volume} {92}},\ \bibinfo {pages} {023003} (\bibinfo {year}
  {2017})}\BibitemShut {NoStop}%
\bibitem [{\citenamefont {Oganessian}\ \emph {et~al.}(2006)\citenamefont
  {Oganessian} \emph {et~al.}}]{Oganessian2006}%
  \BibitemOpen
  \bibfield  {author} {\bibinfo {author} {\bibfnamefont {Y.~T.}\ \bibnamefont
  {Oganessian}} \emph {et~al.},\ }\href {\doibase 10.1103/PhysRevC.74.044602}
  {\bibfield  {journal} {\bibinfo  {journal} {Phys. Rev. C}\ }\textbf {\bibinfo
  {volume} {74}},\ \bibinfo {pages} {044602} (\bibinfo {year}
  {2006})}\BibitemShut {NoStop}%
\bibitem [{\citenamefont {Bender}(2000)}]{Bender2000}%
  \BibitemOpen
  \bibfield  {author} {\bibinfo {author} {\bibfnamefont {M.}~\bibnamefont
  {Bender}},\ }\href {\doibase 10.1103/PhysRevC.61.031302} {\bibfield
  {journal} {\bibinfo  {journal} {Phys. Rev. C}\ }\textbf {\bibinfo {volume}
  {61}},\ \bibinfo {pages} {031302} (\bibinfo {year} {2000})}\BibitemShut
  {NoStop}%
\bibitem [{\citenamefont {Berger}\ \emph {et~al.}(2004)\citenamefont {Berger},
  \citenamefont {Hirata}, \citenamefont {Girod},\ and\ \citenamefont
  {Decharge}}]{Berger2004}%
  \BibitemOpen
  \bibfield  {author} {\bibinfo {author} {\bibfnamefont {J.~F.}\ \bibnamefont
  {Berger}}, \bibinfo {author} {\bibfnamefont {D.}~\bibnamefont {Hirata}},
  \bibinfo {author} {\bibfnamefont {M.}~\bibnamefont {Girod}}, \ and\ \bibinfo
  {author} {\bibfnamefont {J.}~\bibnamefont {Decharge}},\ }\href {\doibase
  10.1142/S021830130400176X} {\bibfield  {journal} {\bibinfo  {journal} {Int.
  J. Mod. Phys. E}\ }\textbf {\bibinfo {volume} {13}},\ \bibinfo {pages} {79}
  (\bibinfo {year} {2004})}\BibitemShut {NoStop}%
\bibitem [{\citenamefont {\ifmmode~\acute{C}\else \'{C}\fi{}wiok}\ \emph
  {et~al.}(1999)\citenamefont {\ifmmode~\acute{C}\else \'{C}\fi{}wiok},
  \citenamefont {Nazarewicz},\ and\ \citenamefont {Heenen}}]{Cwiok1999}%
  \BibitemOpen
  \bibfield  {author} {\bibinfo {author} {\bibfnamefont {S.}~\bibnamefont
  {\ifmmode~\acute{C}\else \'{C}\fi{}wiok}}, \bibinfo {author} {\bibfnamefont
  {W.}~\bibnamefont {Nazarewicz}}, \ and\ \bibinfo {author} {\bibfnamefont
  {P.-H.}\ \bibnamefont {Heenen}},\ }\href {\doibase
  10.1103/PhysRevLett.83.1108} {\bibfield  {journal} {\bibinfo  {journal}
  {Phys. Rev. Lett.}\ }\textbf {\bibinfo {volume} {83}},\ \bibinfo {pages}
  {1108} (\bibinfo {year} {1999})}\BibitemShut {NoStop}%
\bibitem [{\citenamefont {Erler}\ \emph
  {et~al.}(2012{\natexlab{a}})\citenamefont {Erler}, \citenamefont {Langanke},
  \citenamefont {Loens}, \citenamefont {Mart\'{\i}nez-Pinedo},\ and\
  \citenamefont {Reinhard}}]{Erler2012a}%
  \BibitemOpen
  \bibfield  {author} {\bibinfo {author} {\bibfnamefont {J.}~\bibnamefont
  {Erler}}, \bibinfo {author} {\bibfnamefont {K.}~\bibnamefont {Langanke}},
  \bibinfo {author} {\bibfnamefont {H.~P.}\ \bibnamefont {Loens}}, \bibinfo
  {author} {\bibfnamefont {G.}~\bibnamefont {Mart\'{\i}nez-Pinedo}}, \ and\
  \bibinfo {author} {\bibfnamefont {P.-G.}\ \bibnamefont {Reinhard}},\ }\href
  {\doibase 10.1103/PhysRevC.85.025802} {\bibfield  {journal} {\bibinfo
  {journal} {Phys. Rev. C}\ }\textbf {\bibinfo {volume} {85}},\ \bibinfo
  {pages} {025802} (\bibinfo {year} {2012}{\natexlab{a}})}\BibitemShut
  {NoStop}%
\bibitem [{\citenamefont {Gambhir}\ \emph {et~al.}(2005)\citenamefont
  {Gambhir}, \citenamefont {Bhagwat},\ and\ \citenamefont
  {Gupta}}]{Gambhir2005}%
  \BibitemOpen
  \bibfield  {author} {\bibinfo {author} {\bibfnamefont {Y.~K.}\ \bibnamefont
  {Gambhir}}, \bibinfo {author} {\bibfnamefont {A.}~\bibnamefont {Bhagwat}}, \
  and\ \bibinfo {author} {\bibfnamefont {M.}~\bibnamefont {Gupta}},\ }\href
  {\doibase 10.1103/PhysRevC.71.037301} {\bibfield  {journal} {\bibinfo
  {journal} {Phys. Rev. C}\ }\textbf {\bibinfo {volume} {71}},\ \bibinfo
  {pages} {037301} (\bibinfo {year} {2005})}\BibitemShut {NoStop}%
\bibitem [{\citenamefont {Heenen}\ \emph {et~al.}(2015)\citenamefont {Heenen},
  \citenamefont {Skalski}, \citenamefont {Staszczak},\ and\ \citenamefont
  {Vretenar}}]{Heenen2015}%
  \BibitemOpen
  \bibfield  {author} {\bibinfo {author} {\bibfnamefont {P.-H.}\ \bibnamefont
  {Heenen}}, \bibinfo {author} {\bibfnamefont {J.}~\bibnamefont {Skalski}},
  \bibinfo {author} {\bibfnamefont {A.}~\bibnamefont {Staszczak}}, \ and\
  \bibinfo {author} {\bibfnamefont {D.}~\bibnamefont {Vretenar}},\ }\href
  {\doibase https://doi.org/10.1016/j.nuclphysa.2015.07.016} {\bibfield
  {journal} {\bibinfo  {journal} {Nucl. Phys. A}\ }\textbf {\bibinfo {volume}
  {944}},\ \bibinfo {pages} {415 } (\bibinfo {year} {2015})}\BibitemShut
  {NoStop}%
\bibitem [{\citenamefont {Jachimowicz}\ \emph {et~al.}(2014)\citenamefont
  {Jachimowicz}, \citenamefont {Kowal},\ and\ \citenamefont
  {Skalski}}]{Jachimowicz2014}%
  \BibitemOpen
  \bibfield  {author} {\bibinfo {author} {\bibfnamefont {P.}~\bibnamefont
  {Jachimowicz}}, \bibinfo {author} {\bibfnamefont {M.}~\bibnamefont {Kowal}},
  \ and\ \bibinfo {author} {\bibfnamefont {J.}~\bibnamefont {Skalski}},\ }\href
  {\doibase 10.1103/PhysRevC.89.024304} {\bibfield  {journal} {\bibinfo
  {journal} {Phys. Rev. C}\ }\textbf {\bibinfo {volume} {89}},\ \bibinfo
  {pages} {024304} (\bibinfo {year} {2014})}\BibitemShut {NoStop}%
\bibitem [{\citenamefont {Muntian}\ \emph {et~al.}(2003)\citenamefont
  {Muntian}, \citenamefont {Hofmann}, \citenamefont {Patyk},\ and\
  \citenamefont {Sobiczewski}}]{Muntian2003}%
  \BibitemOpen
  \bibfield  {author} {\bibinfo {author} {\bibfnamefont {I.}~\bibnamefont
  {Muntian}}, \bibinfo {author} {\bibfnamefont {S.}~\bibnamefont {Hofmann}},
  \bibinfo {author} {\bibfnamefont {Z.}~\bibnamefont {Patyk}}, \ and\ \bibinfo
  {author} {\bibfnamefont {A.}~\bibnamefont {Sobiczewski}},\ }\href@noop {}
  {\bibfield  {journal} {\bibinfo  {journal} {Acta Phys. Pol.}\ }\textbf
  {\bibinfo {volume} {34}},\ \bibinfo {pages} {2073} (\bibinfo {year}
  {2003})}\BibitemShut {NoStop}%
\bibitem [{\citenamefont {Sobiczewski}\ and\ \citenamefont
  {Pomorski}(2007)}]{Sobiczewski2007}%
  \BibitemOpen
  \bibfield  {author} {\bibinfo {author} {\bibfnamefont {A.}~\bibnamefont
  {Sobiczewski}}\ and\ \bibinfo {author} {\bibfnamefont {K.}~\bibnamefont
  {Pomorski}},\ }\href {\doibase https://doi.org/10.1016/j.ppnp.2006.05.001}
  {\bibfield  {journal} {\bibinfo  {journal} {Prog. Part. Nucl. Phys.}\
  }\textbf {\bibinfo {volume} {58}},\ \bibinfo {pages} {292 } (\bibinfo {year}
  {2007})}\BibitemShut {NoStop}%
\bibitem [{\citenamefont {Tolokonnikov}\ \emph {et~al.}(2013)\citenamefont
  {Tolokonnikov}, \citenamefont {Lutostansky},\ and\ \citenamefont
  {Saperstein}}]{Tolokonnikov2013}%
  \BibitemOpen
  \bibfield  {author} {\bibinfo {author} {\bibfnamefont {S.~V.}\ \bibnamefont
  {Tolokonnikov}}, \bibinfo {author} {\bibfnamefont {Y.~S.}\ \bibnamefont
  {Lutostansky}}, \ and\ \bibinfo {author} {\bibfnamefont {E.~E.}\ \bibnamefont
  {Saperstein}},\ }\href {\doibase 10.1134/S1063778813060136} {\bibfield
  {journal} {\bibinfo  {journal} {Phys. At. Nucl.}\ }\textbf {\bibinfo {volume}
  {76}},\ \bibinfo {pages} {708} (\bibinfo {year} {2013})}\BibitemShut
  {NoStop}%
\bibitem [{\citenamefont {Tolokonnikov}\ \emph {et~al.}(2017)\citenamefont
  {Tolokonnikov}, \citenamefont {Borzov}, \citenamefont {Kortelainen},
  \citenamefont {Lutostansky},\ and\ \citenamefont
  {Saperstein}}]{Tolokonnikov2017}%
  \BibitemOpen
  \bibfield  {author} {\bibinfo {author} {\bibfnamefont {S.~V.}\ \bibnamefont
  {Tolokonnikov}}, \bibinfo {author} {\bibfnamefont {I.~N.}\ \bibnamefont
  {Borzov}}, \bibinfo {author} {\bibfnamefont {M.}~\bibnamefont {Kortelainen}},
  \bibinfo {author} {\bibfnamefont {Y.~S.}\ \bibnamefont {Lutostansky}}, \ and\
  \bibinfo {author} {\bibfnamefont {E.~E.}\ \bibnamefont {Saperstein}},\ }\href
  {\doibase 10.1140/epja/i2017-12220-y} {\bibfield  {journal} {\bibinfo
  {journal} {Eur. Phys. J. A}\ }\textbf {\bibinfo {volume} {53}},\ \bibinfo
  {pages} {33} (\bibinfo {year} {2017})}\BibitemShut {NoStop}%
\bibitem [{\citenamefont {Typel}\ and\ \citenamefont
  {Brown}(2003)}]{Typel2003}%
  \BibitemOpen
  \bibfield  {author} {\bibinfo {author} {\bibfnamefont {S.}~\bibnamefont
  {Typel}}\ and\ \bibinfo {author} {\bibfnamefont {B.~A.}\ \bibnamefont
  {Brown}},\ }\href {\doibase 10.1103/PhysRevC.67.034313} {\bibfield  {journal}
  {\bibinfo  {journal} {Phys. Rev. C}\ }\textbf {\bibinfo {volume} {67}},\
  \bibinfo {pages} {034313} (\bibinfo {year} {2003})}\BibitemShut {NoStop}%
\bibitem [{\citenamefont {Warda}\ and\ \citenamefont
  {Egido}(2012)}]{Warda2012}%
  \BibitemOpen
  \bibfield  {author} {\bibinfo {author} {\bibfnamefont {M.}~\bibnamefont
  {Warda}}\ and\ \bibinfo {author} {\bibfnamefont {J.~L.}\ \bibnamefont
  {Egido}},\ }\href {\doibase 10.1103/PhysRevC.86.014322} {\bibfield  {journal}
  {\bibinfo  {journal} {Phys. Rev. C}\ }\textbf {\bibinfo {volume} {86}},\
  \bibinfo {pages} {014322} (\bibinfo {year} {2012})}\BibitemShut {NoStop}%
\bibitem [{\citenamefont {Brown}(1992)}]{Brown1992}%
  \BibitemOpen
  \bibfield  {author} {\bibinfo {author} {\bibfnamefont {B.~A.}\ \bibnamefont
  {Brown}},\ }\href {\doibase 10.1103/PhysRevC.46.811} {\bibfield  {journal}
  {\bibinfo  {journal} {Phys. Rev. C}\ }\textbf {\bibinfo {volume} {46}},\
  \bibinfo {pages} {811} (\bibinfo {year} {1992})}\BibitemShut {NoStop}%
\bibitem [{\citenamefont {Budaca}\ \emph {et~al.}(2016)\citenamefont {Budaca},
  \citenamefont {Budaca},\ and\ \citenamefont {Silisteanu}}]{Budaca2016}%
  \BibitemOpen
  \bibfield  {author} {\bibinfo {author} {\bibfnamefont {A.}~\bibnamefont
  {Budaca}}, \bibinfo {author} {\bibfnamefont {R.}~\bibnamefont {Budaca}}, \
  and\ \bibinfo {author} {\bibfnamefont {I.}~\bibnamefont {Silisteanu}},\
  }\href {\doibase 10.1016/j.nuclphysa.2016.03.048} {\bibfield  {journal}
  {\bibinfo  {journal} {Nucl. Phys. A}\ }\textbf {\bibinfo {volume} {951}},\
  \bibinfo {pages} {60 } (\bibinfo {year} {2016})}\BibitemShut {NoStop}%
\bibitem [{\citenamefont {Chowdhury}\ \emph {et~al.}(2008)\citenamefont
  {Chowdhury}, \citenamefont {Samanta},\ and\ \citenamefont
  {Basu}}]{Chowdhury2008}%
  \BibitemOpen
  \bibfield  {author} {\bibinfo {author} {\bibfnamefont {P.~R.}\ \bibnamefont
  {Chowdhury}}, \bibinfo {author} {\bibfnamefont {C.}~\bibnamefont {Samanta}},
  \ and\ \bibinfo {author} {\bibfnamefont {D.~N.}\ \bibnamefont {Basu}},\
  }\href {\doibase 10.1103/PhysRevC.77.044603} {\bibfield  {journal} {\bibinfo
  {journal} {Phys. Rev. C}\ }\textbf {\bibinfo {volume} {77}},\ \bibinfo
  {pages} {044603} (\bibinfo {year} {2008})}\BibitemShut {NoStop}%
\bibitem [{\citenamefont {Dong}\ \emph {et~al.}(2011)\citenamefont {Dong},
  \citenamefont {Zuo},\ and\ \citenamefont {Scheid}}]{Dong2011}%
  \BibitemOpen
  \bibfield  {author} {\bibinfo {author} {\bibfnamefont {J.}~\bibnamefont
  {Dong}}, \bibinfo {author} {\bibfnamefont {W.}~\bibnamefont {Zuo}}, \ and\
  \bibinfo {author} {\bibfnamefont {W.}~\bibnamefont {Scheid}},\ }\href
  {\doibase https://doi.org/10.1016/j.nuclphysa.2011.06.016} {\bibfield
  {journal} {\bibinfo  {journal} {Nucl. Phys. A}\ }\textbf {\bibinfo {volume}
  {861}},\ \bibinfo {pages} {1 } (\bibinfo {year} {2011})}\BibitemShut
  {NoStop}%
\bibitem [{\citenamefont {Koura}(2012)}]{Koura2012}%
  \BibitemOpen
  \bibfield  {author} {\bibinfo {author} {\bibfnamefont {H.}~\bibnamefont
  {Koura}},\ }\href {\doibase 10.1080/00223131.2012.704158} {\bibfield
  {journal} {\bibinfo  {journal} {J. Nucl. Sci. Technol.}\ }\textbf {\bibinfo
  {volume} {49}},\ \bibinfo {pages} {816} (\bibinfo {year} {2012})}\BibitemShut
  {NoStop}%
\bibitem [{\citenamefont {Parkhomenko}\ and\ \citenamefont
  {Sobiczewski}(2005)}]{Parkhomenko2005}%
  \BibitemOpen
  \bibfield  {author} {\bibinfo {author} {\bibfnamefont {A.}~\bibnamefont
  {Parkhomenko}}\ and\ \bibinfo {author} {\bibfnamefont {A.}~\bibnamefont
  {Sobiczewski}},\ }\href
  {http://www.actaphys.uj.edu.pl/fulltext?series=Reg&vol=36&page=3095}
  {\bibfield  {journal} {\bibinfo  {journal} {Acta Phys. Pol. B}\ }\textbf
  {\bibinfo {volume} {36}},\ \bibinfo {pages} {3095} (\bibinfo {year}
  {2005})}\BibitemShut {NoStop}%
\bibitem [{\citenamefont {Royer}\ and\ \citenamefont
  {Zhang}(2008)}]{Royer2008}%
  \BibitemOpen
  \bibfield  {author} {\bibinfo {author} {\bibfnamefont {G.}~\bibnamefont
  {Royer}}\ and\ \bibinfo {author} {\bibfnamefont {H.~F.}\ \bibnamefont
  {Zhang}},\ }\href {\doibase 10.1103/PhysRevC.77.037602} {\bibfield  {journal}
  {\bibinfo  {journal} {Phys. Rev. C}\ }\textbf {\bibinfo {volume} {77}},\
  \bibinfo {pages} {037602} (\bibinfo {year} {2008})}\BibitemShut {NoStop}%
\bibitem [{\citenamefont {Ward}\ \emph {et~al.}(2015)\citenamefont {Ward},
  \citenamefont {Carlsson},\ and\ \citenamefont {\AA{}berg}}]{Ward2015}%
  \BibitemOpen
  \bibfield  {author} {\bibinfo {author} {\bibfnamefont {D.~E.}\ \bibnamefont
  {Ward}}, \bibinfo {author} {\bibfnamefont {B.~G.}\ \bibnamefont {Carlsson}},
  \ and\ \bibinfo {author} {\bibfnamefont {S.}~\bibnamefont {\AA{}berg}},\
  }\href {\doibase 10.1103/PhysRevC.92.014314} {\bibfield  {journal} {\bibinfo
  {journal} {Phys. Rev. C}\ }\textbf {\bibinfo {volume} {92}},\ \bibinfo
  {pages} {014314} (\bibinfo {year} {2015})}\BibitemShut {NoStop}%
\bibitem [{\citenamefont {Bender}\ \emph {et~al.}(2003)\citenamefont {Bender},
  \citenamefont {Heenen},\ and\ \citenamefont {Reinhard}}]{Bender2003}%
  \BibitemOpen
  \bibfield  {author} {\bibinfo {author} {\bibfnamefont {M.}~\bibnamefont
  {Bender}}, \bibinfo {author} {\bibfnamefont {P.-H.}\ \bibnamefont {Heenen}},
  \ and\ \bibinfo {author} {\bibfnamefont {P.-G.}\ \bibnamefont {Reinhard}},\
  }\href {\doibase 10.1103/RevModPhys.75.121} {\bibfield  {journal} {\bibinfo
  {journal} {Rev. Mod. Phys.}\ }\textbf {\bibinfo {volume} {75}},\ \bibinfo
  {pages} {121} (\bibinfo {year} {2003})}\BibitemShut {NoStop}%
\bibitem [{\citenamefont {Stoitsov}\ \emph {et~al.}(2006)\citenamefont
  {Stoitsov}, \citenamefont {Dobaczewski}, \citenamefont {Nazarewicz},\ and\
  \citenamefont {Borycki}}]{Stoitsov2006}%
  \BibitemOpen
  \bibfield  {author} {\bibinfo {author} {\bibfnamefont {M.}~\bibnamefont
  {Stoitsov}}, \bibinfo {author} {\bibfnamefont {J.}~\bibnamefont
  {Dobaczewski}}, \bibinfo {author} {\bibfnamefont {W.}~\bibnamefont
  {Nazarewicz}}, \ and\ \bibinfo {author} {\bibfnamefont {P.}~\bibnamefont
  {Borycki}},\ }\href {\doibase https://doi.org/10.1016/j.ijms.2006.01.040}
  {\bibfield  {journal} {\bibinfo  {journal} {Int. J. Mass Spectrom.}\ }\textbf
  {\bibinfo {volume} {251}},\ \bibinfo {pages} {243 } (\bibinfo {year}
  {2006})}\BibitemShut {NoStop}%
\bibitem [{\citenamefont {Kl\"upfel}\ \emph {et~al.}(2009)\citenamefont
  {Kl\"upfel}, \citenamefont {Reinhard}, \citenamefont {B\"urvenich},\ and\
  \citenamefont {Maruhn}}]{Klupfel2009}%
  \BibitemOpen
  \bibfield  {author} {\bibinfo {author} {\bibfnamefont {P.}~\bibnamefont
  {Kl\"upfel}}, \bibinfo {author} {\bibfnamefont {P.-G.}\ \bibnamefont
  {Reinhard}}, \bibinfo {author} {\bibfnamefont {T.~J.}\ \bibnamefont
  {B\"urvenich}}, \ and\ \bibinfo {author} {\bibfnamefont {J.~A.}\ \bibnamefont
  {Maruhn}},\ }\href {\doibase 10.1103/PhysRevC.79.034310} {\bibfield
  {journal} {\bibinfo  {journal} {Phys. Rev. C}\ }\textbf {\bibinfo {volume}
  {79}},\ \bibinfo {pages} {034310} (\bibinfo {year} {2009})}\BibitemShut
  {NoStop}%
\bibitem [{\citenamefont {Kortelainen}\ \emph {et~al.}(2010)\citenamefont
  {Kortelainen}, \citenamefont {Lesinski}, \citenamefont {Mor\'e},
  \citenamefont {Nazarewicz}, \citenamefont {Sarich}, \citenamefont {Schunck},
  \citenamefont {Stoitsov},\ and\ \citenamefont {Wild}}]{Kortelainen2010}%
  \BibitemOpen
  \bibfield  {author} {\bibinfo {author} {\bibfnamefont {M.}~\bibnamefont
  {Kortelainen}}, \bibinfo {author} {\bibfnamefont {T.}~\bibnamefont
  {Lesinski}}, \bibinfo {author} {\bibfnamefont {J.}~\bibnamefont {Mor\'e}},
  \bibinfo {author} {\bibfnamefont {W.}~\bibnamefont {Nazarewicz}}, \bibinfo
  {author} {\bibfnamefont {J.}~\bibnamefont {Sarich}}, \bibinfo {author}
  {\bibfnamefont {N.}~\bibnamefont {Schunck}}, \bibinfo {author} {\bibfnamefont
  {M.~V.}\ \bibnamefont {Stoitsov}}, \ and\ \bibinfo {author} {\bibfnamefont
  {S.}~\bibnamefont {Wild}},\ }\href {\doibase 10.1103/PhysRevC.82.024313}
  {\bibfield  {journal} {\bibinfo  {journal} {Phys. Rev. C}\ }\textbf {\bibinfo
  {volume} {82}},\ \bibinfo {pages} {024313} (\bibinfo {year}
  {2010})}\BibitemShut {NoStop}%
\bibitem [{\citenamefont {Skyrme}(1958)}]{Skyrme1958}%
  \BibitemOpen
  \bibfield  {author} {\bibinfo {author} {\bibfnamefont {T.}~\bibnamefont
  {Skyrme}},\ }\href {\doibase https://doi.org/10.1016/0029-5582(58)90345-6}
  {\bibfield  {journal} {\bibinfo  {journal} {Nucl. Phys.}\ }\textbf {\bibinfo
  {volume} {9}},\ \bibinfo {pages} {615 } (\bibinfo {year} {1958})}\BibitemShut
  {NoStop}%
\bibitem [{\citenamefont {Vautherin}\ and\ \citenamefont
  {Brink}(1972)}]{Vautherin1972}%
  \BibitemOpen
  \bibfield  {author} {\bibinfo {author} {\bibfnamefont {D.}~\bibnamefont
  {Vautherin}}\ and\ \bibinfo {author} {\bibfnamefont {D.~M.}\ \bibnamefont
  {Brink}},\ }\href {\doibase 10.1103/PhysRevC.5.626} {\bibfield  {journal}
  {\bibinfo  {journal} {Phys. Rev. C}\ }\textbf {\bibinfo {volume} {5}},\
  \bibinfo {pages} {626} (\bibinfo {year} {1972})}\BibitemShut {NoStop}%
\bibitem [{\citenamefont {Dobaczewski}\ \emph {et~al.}(2002)\citenamefont
  {Dobaczewski}, \citenamefont {Nazarewicz},\ and\ \citenamefont
  {Stoitsov}}]{Dobaczewski2002}%
  \BibitemOpen
  \bibfield  {author} {\bibinfo {author} {\bibfnamefont {J.}~\bibnamefont
  {Dobaczewski}}, \bibinfo {author} {\bibfnamefont {W.}~\bibnamefont
  {Nazarewicz}}, \ and\ \bibinfo {author} {\bibfnamefont {M.}~\bibnamefont
  {Stoitsov}},\ }\href {\doibase 10.1140/epja/i2001-10218-8} {\bibfield
  {journal} {\bibinfo  {journal} {Eur. Phys. J. A}\ }\textbf {\bibinfo {volume}
  {15}},\ \bibinfo {pages} {21} (\bibinfo {year} {2002})}\BibitemShut {NoStop}%
\bibitem [{\citenamefont {Bartel}\ \emph {et~al.}(1982)\citenamefont {Bartel},
  \citenamefont {Quentin}, \citenamefont {Brack}, \citenamefont {Guet},\ and\
  \citenamefont {H\r{a}kansson}}]{Bartel1982}%
  \BibitemOpen
  \bibfield  {author} {\bibinfo {author} {\bibfnamefont {J.}~\bibnamefont
  {Bartel}}, \bibinfo {author} {\bibfnamefont {P.}~\bibnamefont {Quentin}},
  \bibinfo {author} {\bibfnamefont {M.}~\bibnamefont {Brack}}, \bibinfo
  {author} {\bibfnamefont {C.}~\bibnamefont {Guet}}, \ and\ \bibinfo {author}
  {\bibfnamefont {H.-B.}\ \bibnamefont {H\r{a}kansson}},\ }\href {\doibase
  http://dx.doi.org/10.1016/0375-9474(82)90403-1} {\bibfield  {journal}
  {\bibinfo  {journal} {Nucl. Phys. A}\ }\textbf {\bibinfo {volume} {386}},\
  \bibinfo {pages} {79 } (\bibinfo {year} {1982})}\BibitemShut {NoStop}%
\bibitem [{\citenamefont {Chabanat}\ \emph {et~al.}(1998)\citenamefont
  {Chabanat}, \citenamefont {Bonche}, \citenamefont {Haensel}, \citenamefont
  {Meyer},\ and\ \citenamefont {Schaeffer}}]{Chabanat1998}%
  \BibitemOpen
  \bibfield  {author} {\bibinfo {author} {\bibfnamefont {E.}~\bibnamefont
  {Chabanat}}, \bibinfo {author} {\bibfnamefont {P.}~\bibnamefont {Bonche}},
  \bibinfo {author} {\bibfnamefont {P.}~\bibnamefont {Haensel}}, \bibinfo
  {author} {\bibfnamefont {J.}~\bibnamefont {Meyer}}, \ and\ \bibinfo {author}
  {\bibfnamefont {R.}~\bibnamefont {Schaeffer}},\ }\href {\doibase
  http://dx.doi.org/10.1016/S0375-9474(98)00180-8} {\bibfield  {journal}
  {\bibinfo  {journal} {Nucl. Phys. A}\ }\textbf {\bibinfo {volume} {635}},\
  \bibinfo {pages} {231 } (\bibinfo {year} {1998})}\BibitemShut {NoStop}%
\bibitem [{\citenamefont {Kortelainen}\ \emph {et~al.}(2012)\citenamefont
  {Kortelainen}, \citenamefont {McDonnell}, \citenamefont {Nazarewicz},
  \citenamefont {Reinhard}, \citenamefont {Sarich}, \citenamefont {Schunck},
  \citenamefont {Stoitsov},\ and\ \citenamefont {Wild}}]{Kortelainen2012}%
  \BibitemOpen
  \bibfield  {author} {\bibinfo {author} {\bibfnamefont {M.}~\bibnamefont
  {Kortelainen}}, \bibinfo {author} {\bibfnamefont {J.}~\bibnamefont
  {McDonnell}}, \bibinfo {author} {\bibfnamefont {W.}~\bibnamefont
  {Nazarewicz}}, \bibinfo {author} {\bibfnamefont {P.-G.}\ \bibnamefont
  {Reinhard}}, \bibinfo {author} {\bibfnamefont {J.}~\bibnamefont {Sarich}},
  \bibinfo {author} {\bibfnamefont {N.}~\bibnamefont {Schunck}}, \bibinfo
  {author} {\bibfnamefont {M.~V.}\ \bibnamefont {Stoitsov}}, \ and\ \bibinfo
  {author} {\bibfnamefont {S.~M.}\ \bibnamefont {Wild}},\ }\href {\doibase
  10.1103/PhysRevC.85.024304} {\bibfield  {journal} {\bibinfo  {journal} {Phys.
  Rev. C}\ }\textbf {\bibinfo {volume} {85}},\ \bibinfo {pages} {024304}
  (\bibinfo {year} {2012})}\BibitemShut {NoStop}%
\bibitem [{\citenamefont {Kortelainen}\ \emph {et~al.}(2014)\citenamefont
  {Kortelainen}, \citenamefont {McDonnell}, \citenamefont {Nazarewicz},
  \citenamefont {Olsen}, \citenamefont {Reinhard}, \citenamefont {Sarich},
  \citenamefont {Schunck}, \citenamefont {Wild}, \citenamefont {Davesne},
  \citenamefont {Erler},\ and\ \citenamefont {Pastore}}]{Kortelainen2014}%
  \BibitemOpen
  \bibfield  {author} {\bibinfo {author} {\bibfnamefont {M.}~\bibnamefont
  {Kortelainen}}, \bibinfo {author} {\bibfnamefont {J.}~\bibnamefont
  {McDonnell}}, \bibinfo {author} {\bibfnamefont {W.}~\bibnamefont
  {Nazarewicz}}, \bibinfo {author} {\bibfnamefont {E.}~\bibnamefont {Olsen}},
  \bibinfo {author} {\bibfnamefont {P.-G.}\ \bibnamefont {Reinhard}}, \bibinfo
  {author} {\bibfnamefont {J.}~\bibnamefont {Sarich}}, \bibinfo {author}
  {\bibfnamefont {N.}~\bibnamefont {Schunck}}, \bibinfo {author} {\bibfnamefont
  {S.~M.}\ \bibnamefont {Wild}}, \bibinfo {author} {\bibfnamefont
  {D.}~\bibnamefont {Davesne}}, \bibinfo {author} {\bibfnamefont
  {J.}~\bibnamefont {Erler}}, \ and\ \bibinfo {author} {\bibfnamefont
  {A.}~\bibnamefont {Pastore}},\ }\href {\doibase 10.1103/PhysRevC.89.054314}
  {\bibfield  {journal} {\bibinfo  {journal} {Phys. Rev. C}\ }\textbf {\bibinfo
  {volume} {89}},\ \bibinfo {pages} {054314} (\bibinfo {year}
  {2014})}\BibitemShut {NoStop}%
\bibitem [{\citenamefont {Shi}\ \emph {et~al.}(2014)\citenamefont {Shi},
  \citenamefont {Dobaczewski},\ and\ \citenamefont {Greenlees}}]{Shi2014}%
  \BibitemOpen
  \bibfield  {author} {\bibinfo {author} {\bibfnamefont {Y.}~\bibnamefont
  {Shi}}, \bibinfo {author} {\bibfnamefont {J.}~\bibnamefont {Dobaczewski}}, \
  and\ \bibinfo {author} {\bibfnamefont {P.~T.}\ \bibnamefont {Greenlees}},\
  }\href {\doibase 10.1103/PhysRevC.89.034309} {\bibfield  {journal} {\bibinfo
  {journal} {Phys. Rev. C}\ }\textbf {\bibinfo {volume} {89}},\ \bibinfo
  {pages} {034309} (\bibinfo {year} {2014})}\BibitemShut {NoStop}%
\bibitem [{\citenamefont {Erler}\ \emph
  {et~al.}(2012{\natexlab{b}})\citenamefont {Erler}, \citenamefont {Birge},
  \citenamefont {Kortelainen}, \citenamefont {Nazarewicz}, \citenamefont
  {Olsen}, \citenamefont {Perhac},\ and\ \citenamefont
  {Stoitsov}}]{Erler2012b}%
  \BibitemOpen
  \bibfield  {author} {\bibinfo {author} {\bibfnamefont {J.}~\bibnamefont
  {Erler}}, \bibinfo {author} {\bibfnamefont {N.}~\bibnamefont {Birge}},
  \bibinfo {author} {\bibfnamefont {M.}~\bibnamefont {Kortelainen}}, \bibinfo
  {author} {\bibfnamefont {W.}~\bibnamefont {Nazarewicz}}, \bibinfo {author}
  {\bibfnamefont {E.}~\bibnamefont {Olsen}}, \bibinfo {author} {\bibfnamefont
  {A.~M.}\ \bibnamefont {Perhac}}, \ and\ \bibinfo {author} {\bibfnamefont
  {M.}~\bibnamefont {Stoitsov}},\ }\href
  {http://dx.doi.org/10.1038/nature11188} {\bibfield  {journal} {\bibinfo
  {journal} {Nature}\ }\textbf {\bibinfo {volume} {486}},\ \bibinfo {pages}
  {509} (\bibinfo {year} {2012}{\natexlab{b}})}\BibitemShut {NoStop}%
\bibitem [{\citenamefont {Ring}\ and\ \citenamefont {Schuck}(1980)}]{Ring1980}%
  \BibitemOpen
  \bibfield  {author} {\bibinfo {author} {\bibfnamefont {P.}~\bibnamefont
  {Ring}}\ and\ \bibinfo {author} {\bibfnamefont {P.}~\bibnamefont {Schuck}},\
  }\href@noop {} {\emph {\bibinfo {title} {The Nuclear Many-Body Problem}}}\
  (\bibinfo  {publisher} {Springer},\ \bibinfo {address} {New York},\ \bibinfo
  {year} {1980})\BibitemShut {NoStop}%
\bibitem [{\citenamefont {Afanasjev}(2015)}]{Afanasjev2015}%
  \BibitemOpen
  \bibfield  {author} {\bibinfo {author} {\bibfnamefont {A.~V.}\ \bibnamefont
  {Afanasjev}},\ }\href {http://stacks.iop.org/0954-3899/42/i=3/a=034002}
  {\bibfield  {journal} {\bibinfo  {journal} {J. Phys. G.}\ }\textbf {\bibinfo
  {volume} {42}},\ \bibinfo {pages} {034002} (\bibinfo {year}
  {2015})}\BibitemShut {NoStop}%
\bibitem [{\citenamefont {Bonneau}\ \emph {et~al.}(2007)\citenamefont
  {Bonneau}, \citenamefont {Quentin},\ and\ \citenamefont
  {M\"oller}}]{Bonneau2007}%
  \BibitemOpen
  \bibfield  {author} {\bibinfo {author} {\bibfnamefont {L.}~\bibnamefont
  {Bonneau}}, \bibinfo {author} {\bibfnamefont {P.}~\bibnamefont {Quentin}}, \
  and\ \bibinfo {author} {\bibfnamefont {P.}~\bibnamefont {M\"oller}},\ }\href
  {\doibase 10.1103/PhysRevC.76.024320} {\bibfield  {journal} {\bibinfo
  {journal} {Phys. Rev. C}\ }\textbf {\bibinfo {volume} {76}},\ \bibinfo
  {pages} {024320} (\bibinfo {year} {2007})}\BibitemShut {NoStop}%
\bibitem [{\citenamefont {Schunck}\ \emph {et~al.}(2010)\citenamefont
  {Schunck}, \citenamefont {Dobaczewski}, \citenamefont {McDonnell},
  \citenamefont {Mor\'e}, \citenamefont {Nazarewicz}, \citenamefont {Sarich},\
  and\ \citenamefont {Stoitsov}}]{Schunck2010}%
  \BibitemOpen
  \bibfield  {author} {\bibinfo {author} {\bibfnamefont {N.}~\bibnamefont
  {Schunck}}, \bibinfo {author} {\bibfnamefont {J.}~\bibnamefont
  {Dobaczewski}}, \bibinfo {author} {\bibfnamefont {J.}~\bibnamefont
  {McDonnell}}, \bibinfo {author} {\bibfnamefont {J.}~\bibnamefont {Mor\'e}},
  \bibinfo {author} {\bibfnamefont {W.}~\bibnamefont {Nazarewicz}}, \bibinfo
  {author} {\bibfnamefont {J.}~\bibnamefont {Sarich}}, \ and\ \bibinfo {author}
  {\bibfnamefont {M.~V.}\ \bibnamefont {Stoitsov}},\ }\href {\doibase
  10.1103/PhysRevC.81.024316} {\bibfield  {journal} {\bibinfo  {journal} {Phys.
  Rev. C}\ }\textbf {\bibinfo {volume} {81}},\ \bibinfo {pages} {024316}
  (\bibinfo {year} {2010})}\BibitemShut {NoStop}%
\bibitem [{\citenamefont {Perez}\ \emph {et~al.}(2017)\citenamefont {Perez},
  \citenamefont {Schunck}, \citenamefont {Lasseri}, \citenamefont {Zhang},\
  and\ \citenamefont {Sarich}}]{Navarro2017}%
  \BibitemOpen
  \bibfield  {author} {\bibinfo {author} {\bibfnamefont {R.~N.}\ \bibnamefont
  {Perez}}, \bibinfo {author} {\bibfnamefont {N.}~\bibnamefont {Schunck}},
  \bibinfo {author} {\bibfnamefont {R.-D.}\ \bibnamefont {Lasseri}}, \bibinfo
  {author} {\bibfnamefont {C.}~\bibnamefont {Zhang}}, \ and\ \bibinfo {author}
  {\bibfnamefont {J.}~\bibnamefont {Sarich}},\ }\href {\doibase
  https://doi.org/10.1016/j.cpc.2017.06.022} {\bibfield  {journal} {\bibinfo
  {journal} {Comp. Phys. Comm.}\ }\textbf {\bibinfo {volume} {220}},\ \bibinfo
  {pages} {363 } (\bibinfo {year} {2017})}\BibitemShut {NoStop}%
\bibitem [{\citenamefont {{\'C}wiok}\ \emph {et~al.}(2005)\citenamefont
  {{\'C}wiok}, \citenamefont {Heenen},\ and\ \citenamefont
  {Nazarewicz}}]{Cwiok2005}%
  \BibitemOpen
  \bibfield  {author} {\bibinfo {author} {\bibfnamefont {S.}~\bibnamefont
  {{\'C}wiok}}, \bibinfo {author} {\bibfnamefont {P.-H.}\ \bibnamefont
  {Heenen}}, \ and\ \bibinfo {author} {\bibfnamefont {W.}~\bibnamefont
  {Nazarewicz}},\ }\href {http://dx.doi.org/10.1038/nature03336} {\bibfield
  {journal} {\bibinfo  {journal} {Nature}\ }\textbf {\bibinfo {volume} {433}},\
  \bibinfo {pages} {705} (\bibinfo {year} {2005})},\ \bibinfo {note} {review
  Article}\BibitemShut {NoStop}%
\bibitem [{\citenamefont {M{\"o}ller}\ \emph {et~al.}(2008)\citenamefont
  {M{\"o}ller}, \citenamefont {Bengtsson}, \citenamefont {Carlsson},
  \citenamefont {Olivius}, \citenamefont {Ichikawa}, \citenamefont {Sagawa},\
  and\ \citenamefont {Iwamoto}}]{Moller2008}%
  \BibitemOpen
  \bibfield  {author} {\bibinfo {author} {\bibfnamefont {P.}~\bibnamefont
  {M{\"o}ller}}, \bibinfo {author} {\bibfnamefont {R.}~\bibnamefont
  {Bengtsson}}, \bibinfo {author} {\bibfnamefont {B.}~\bibnamefont {Carlsson}},
  \bibinfo {author} {\bibfnamefont {P.}~\bibnamefont {Olivius}}, \bibinfo
  {author} {\bibfnamefont {T.}~\bibnamefont {Ichikawa}}, \bibinfo {author}
  {\bibfnamefont {H.}~\bibnamefont {Sagawa}}, \ and\ \bibinfo {author}
  {\bibfnamefont {A.}~\bibnamefont {Iwamoto}},\ }\href {\doibase
  https://doi.org/10.1016/j.adt.2008.05.002} {\bibfield  {journal} {\bibinfo
  {journal} {At. Data Nucl. Data Tables}\ }\textbf {\bibinfo {volume} {94}},\
  \bibinfo {pages} {758 } (\bibinfo {year} {2008})}\BibitemShut {NoStop}%
\bibitem [{\citenamefont {Stoitsov}\ \emph {et~al.}(2007)\citenamefont
  {Stoitsov}, \citenamefont {Dobaczewski}, \citenamefont {Kirchner},
  \citenamefont {Nazarewicz},\ and\ \citenamefont {Terasaki}}]{Stoitsov2007}%
  \BibitemOpen
  \bibfield  {author} {\bibinfo {author} {\bibfnamefont {M.~V.}\ \bibnamefont
  {Stoitsov}}, \bibinfo {author} {\bibfnamefont {J.}~\bibnamefont
  {Dobaczewski}}, \bibinfo {author} {\bibfnamefont {R.}~\bibnamefont
  {Kirchner}}, \bibinfo {author} {\bibfnamefont {W.}~\bibnamefont
  {Nazarewicz}}, \ and\ \bibinfo {author} {\bibfnamefont {J.}~\bibnamefont
  {Terasaki}},\ }\href {\doibase 10.1103/PhysRevC.76.014308} {\bibfield
  {journal} {\bibinfo  {journal} {Phys. Rev. C}\ }\textbf {\bibinfo {volume}
  {76}},\ \bibinfo {pages} {014308} (\bibinfo {year} {2007})}\BibitemShut
  {NoStop}%
\bibitem [{mas()}]{massexplorer}%
  \BibitemOpen
  \href@noop {} {}\bibinfo {howpublished}
  {\url{http://massexplorer.frib.msu.edu}}\BibitemShut {NoStop}%
\bibitem [{\citenamefont {Wang}\ \emph {et~al.}(2017)\citenamefont {Wang},
  \citenamefont {Audi}, \citenamefont {Kondev}, \citenamefont {Huang},
  \citenamefont {Naimi},\ and\ \citenamefont {Xu}}]{Wang2017}%
  \BibitemOpen
  \bibfield  {author} {\bibinfo {author} {\bibfnamefont {M.}~\bibnamefont
  {Wang}}, \bibinfo {author} {\bibfnamefont {G.}~\bibnamefont {Audi}}, \bibinfo
  {author} {\bibfnamefont {F.}~\bibnamefont {Kondev}}, \bibinfo {author}
  {\bibfnamefont {W.}~\bibnamefont {Huang}}, \bibinfo {author} {\bibfnamefont
  {S.}~\bibnamefont {Naimi}}, \ and\ \bibinfo {author} {\bibfnamefont
  {X.}~\bibnamefont {Xu}},\ }\href
  {http://stacks.iop.org/1674-1137/41/i=3/a=030003} {\bibfield  {journal}
  {\bibinfo  {journal} {Chin. Phys. C}\ }\textbf {\bibinfo {volume} {41}},\
  \bibinfo {pages} {030003} (\bibinfo {year} {2017})}\BibitemShut {NoStop}%
\bibitem [{\citenamefont {Brewer}\ \emph {et~al.}(2018)\citenamefont {Brewer}
  \emph {et~al.}}]{Brewer2018}%
  \BibitemOpen
  \bibfield  {author} {\bibinfo {author} {\bibfnamefont {N.~T.}\ \bibnamefont
  {Brewer}} \emph {et~al.},\ }\href {\doibase 10.1103/PhysRevC.98.024317}
  {\bibfield  {journal} {\bibinfo  {journal} {Phys. Rev. C}\ }\textbf {\bibinfo
  {volume} {98}},\ \bibinfo {pages} {024317} (\bibinfo {year}
  {2018})}\BibitemShut {NoStop}%
\bibitem [{\citenamefont {{\'C}wiok}\ \emph {et~al.}(1983)\citenamefont
  {{\'C}wiok}, \citenamefont {Pashkevich}, \citenamefont {Dudek},\ and\
  \citenamefont {Nazarewicz}}]{Cwiok1983}%
  \BibitemOpen
  \bibfield  {author} {\bibinfo {author} {\bibfnamefont {S.}~\bibnamefont
  {{\'C}wiok}}, \bibinfo {author} {\bibfnamefont {V.}~\bibnamefont
  {Pashkevich}}, \bibinfo {author} {\bibfnamefont {J.}~\bibnamefont {Dudek}}, \
  and\ \bibinfo {author} {\bibfnamefont {W.}~\bibnamefont {Nazarewicz}},\
  }\href {\doibase https://doi.org/10.1016/0375-9474(83)90201-4} {\bibfield
  {journal} {\bibinfo  {journal} {Nucl. Phys. A}\ }\textbf {\bibinfo {volume}
  {410}},\ \bibinfo {pages} {254 } (\bibinfo {year} {1983})}\BibitemShut
  {NoStop}%
\bibitem [{\citenamefont {M{\"o}ller}\ and\ \citenamefont
  {Nix}(1994)}]{Moller1994}%
  \BibitemOpen
  \bibfield  {author} {\bibinfo {author} {\bibfnamefont {P.}~\bibnamefont
  {M{\"o}ller}}\ and\ \bibinfo {author} {\bibfnamefont {J.~R.}\ \bibnamefont
  {Nix}},\ }\href {http://stacks.iop.org/0954-3899/20/i=11/a=003} {\bibfield
  {journal} {\bibinfo  {journal} {J. Phys. G}\ }\textbf {\bibinfo {volume}
  {20}},\ \bibinfo {pages} {1681} (\bibinfo {year} {1994})}\BibitemShut
  {NoStop}%
\bibitem [{\citenamefont {Bernardo}\ and\ \citenamefont
  {Smith}(1994)}]{Bernardo1994}%
  \BibitemOpen
  \bibfield  {author} {\bibinfo {author} {\bibfnamefont {J.~M.}\ \bibnamefont
  {Bernardo}}\ and\ \bibinfo {author} {\bibfnamefont {A.~F.~M.}\ \bibnamefont
  {Smith}},\ }\href@noop {} {\emph {\bibinfo {title} {Bayesian Theory}}}\
  (\bibinfo  {publisher} {Wiley},\ \bibinfo {address} {New Jersey},\ \bibinfo
  {year} {1994})\BibitemShut {NoStop}%
\bibitem [{\citenamefont {Gates}\ \emph {et~al.}(2015)\citenamefont {Gates}
  \emph {et~al.}}]{Gates2015}%
  \BibitemOpen
  \bibfield  {author} {\bibinfo {author} {\bibfnamefont {J.~M.}\ \bibnamefont
  {Gates}} \emph {et~al.},\ }\href {\doibase 10.1103/PhysRevC.92.021301}
  {\bibfield  {journal} {\bibinfo  {journal} {Phys. Rev. C}\ }\textbf {\bibinfo
  {volume} {92}},\ \bibinfo {pages} {021301} (\bibinfo {year}
  {2015})}\BibitemShut {NoStop}%
\bibitem [{\citenamefont {Gates}(2016)}]{Gates2016}%
  \BibitemOpen
  \bibfield  {author} {\bibinfo {author} {\bibfnamefont {J.~M.}\ \bibnamefont
  {Gates}},\ }\href {\doibase 10.1051/epjconf/201613108003} {\bibfield
  {journal} {\bibinfo  {journal} {EPJ Web Conf.}\ }\textbf {\bibinfo {volume}
  {131}},\ \bibinfo {pages} {08003} (\bibinfo {year} {2016})}\BibitemShut
  {NoStop}%
\bibitem [{\citenamefont {Gates}\ \emph {et~al.}(2018)\citenamefont {Gates}
  \emph {et~al.}}]{Gates2018}%
  \BibitemOpen
  \bibfield  {author} {\bibinfo {author} {\bibfnamefont {J.~M.}\ \bibnamefont
  {Gates}} \emph {et~al.},\ }\href
  {https://journals.aps.org/prl/accepted/10075Yc7Gd61c46a65162be0786515991684e13b8}
  {\bibfield  {journal} {\bibinfo  {journal} {Phys. Rev. Lett.}\ } (\bibinfo
  {year} {2018})}\BibitemShut {NoStop}%
\bibitem [{\citenamefont {D{\"u}llmann}(2016)}]{Dullmann2016}%
  \BibitemOpen
  \bibfield  {author} {\bibinfo {author} {\bibfnamefont {C.~E.}\ \bibnamefont
  {D{\"u}llmann}},\ }\href {\doibase 10.1051/epjconf/201613108004} {\bibfield
  {journal} {\bibinfo  {journal} {EPJ Web Conf.}\ }\textbf {\bibinfo {volume}
  {131}},\ \bibinfo {pages} {08004} (\bibinfo {year} {2016})}\BibitemShut
  {NoStop}%
\bibitem [{\citenamefont {He{\ss}berger}\ and\ \citenamefont
  {Ackermann}(2017)}]{Hessberger2017}%
  \BibitemOpen
  \bibfield  {author} {\bibinfo {author} {\bibfnamefont {F.~P.}\ \bibnamefont
  {He{\ss}berger}}\ and\ \bibinfo {author} {\bibfnamefont {D.}~\bibnamefont
  {Ackermann}},\ }\href {\doibase 10.1140/epja/i2017-12307-5} {\bibfield
  {journal} {\bibinfo  {journal} {Eur. Phys. J. A}\ }\textbf {\bibinfo {volume}
  {53}},\ \bibinfo {pages} {123} (\bibinfo {year} {2017})}\BibitemShut
  {NoStop}%
\bibitem [{\citenamefont {Hofmann}\ \emph {et~al.}(2016)\citenamefont {Hofmann}
  \emph {et~al.}}]{Hofmann2016}%
  \BibitemOpen
  \bibfield  {author} {\bibinfo {author} {\bibfnamefont {S.}~\bibnamefont
  {Hofmann}} \emph {et~al.},\ }\href {\doibase 10.1140/epja/i2016-16180-4}
  {\bibfield  {journal} {\bibinfo  {journal} {Eur. Phys. J. A}\ }\textbf
  {\bibinfo {volume} {52}},\ \bibinfo {pages} {180} (\bibinfo {year}
  {2016})}\BibitemShut {NoStop}%
\bibitem [{\citenamefont {Roberto}\ and\ \citenamefont
  {Rykaczewski}(2018)}]{Roberto2018}%
  \BibitemOpen
  \bibfield  {author} {\bibinfo {author} {\bibfnamefont {J.~B.}\ \bibnamefont
  {Roberto}}\ and\ \bibinfo {author} {\bibfnamefont {K.~P.}\ \bibnamefont
  {Rykaczewski}},\ }\href {\doibase 10.1080/01496395.2017.1290658} {\bibfield
  {journal} {\bibinfo  {journal} {Sep. Sci. Technol.}\ }\textbf {\bibinfo
  {volume} {53}},\ \bibinfo {pages} {1813} (\bibinfo {year}
  {2018})}\BibitemShut {NoStop}%
\bibitem [{\citenamefont {Hoeting}\ \emph {et~al.}(1999)\citenamefont
  {Hoeting}, \citenamefont {Madigan}, \citenamefont {Raftery},\ and\
  \citenamefont {Volinsky}}]{Hoeting1999}%
  \BibitemOpen
  \bibfield  {author} {\bibinfo {author} {\bibfnamefont {J.~A.}\ \bibnamefont
  {Hoeting}}, \bibinfo {author} {\bibfnamefont {D.}~\bibnamefont {Madigan}},
  \bibinfo {author} {\bibfnamefont {A.~E.}\ \bibnamefont {Raftery}}, \ and\
  \bibinfo {author} {\bibfnamefont {C.~T.}\ \bibnamefont {Volinsky}},\ }\href
  {\doibase 10.1214/ss/1009212519} {\bibfield  {journal} {\bibinfo  {journal}
  {Statist. Sci.}\ }\textbf {\bibinfo {volume} {14}},\ \bibinfo {pages} {382}
  (\bibinfo {year} {1999})}\BibitemShut {NoStop}%
\bibitem [{\citenamefont {Wasserman}(2000)}]{Wasserman2000}%
  \BibitemOpen
  \bibfield  {author} {\bibinfo {author} {\bibfnamefont {L.}~\bibnamefont
  {Wasserman}},\ }\href {\doibase https://doi.org/10.1006/jmps.1999.1278}
  {\bibfield  {journal} {\bibinfo  {journal} {J. Math. Psych.}\ }\textbf
  {\bibinfo {volume} {44}},\ \bibinfo {pages} {92 } (\bibinfo {year}
  {2000})}\BibitemShut {NoStop}%
\bibitem [{\citenamefont {Neufcourt}\ \emph {et~al.}(2018)\citenamefont
  {Neufcourt}, \citenamefont {Cao}, \citenamefont {Nazarewicz},\ and\
  \citenamefont {Viens}}]{Neufcourt2018}%
  \BibitemOpen
  \bibfield  {author} {\bibinfo {author} {\bibfnamefont {L.}~\bibnamefont
  {Neufcourt}}, \bibinfo {author} {\bibfnamefont {Y.}~\bibnamefont {Cao}},
  \bibinfo {author} {\bibfnamefont {W.}~\bibnamefont {Nazarewicz}}, \ and\
  \bibinfo {author} {\bibfnamefont {F.}~\bibnamefont {Viens}},\ }\href
  {\doibase 10.1103/PhysRevC.98.034318} {\bibfield  {journal} {\bibinfo
  {journal} {Phys. Rev. C}\ }\textbf {\bibinfo {volume} {98}},\ \bibinfo
  {pages} {034318} (\bibinfo {year} {2018})}\BibitemShut {NoStop}%
\end{thebibliography}%

\end{document}